\documentclass[
  aps,prd,            % 明确期刊
  reprint,            % 贴近期刊排版（PRD 默认双栏）
  superscriptaddress,
  nofootinbib,
  amsmath,amssymb
]{revtex4-2}

\usepackage[utf8]{inputenc}

\usepackage{bm}
\usepackage{dcolumn}
\usepackage{graphicx}
\graphicspath{{figure/}}

\usepackage[usenames,dvipsnames,svgnames]{xcolor}

\usepackage[normalem]{ulem}  % 关键：normalem
\usepackage{verbatim}
\usepackage{makecell}
\usepackage{xspace}

\usepackage{hyperref}
\usepackage[all]{hypcap}

\definecolor{darkgreen}{cmyk}{0.9,0,0.9,0.6}
\definecolor{brickred}{RGB}{210,65,84}
\hypersetup{
  colorlinks=true,
  citecolor=Blue,
  linkcolor=Magenta,
  filecolor=NavyBlue,
  urlcolor=NavyBlue
}

\newcommand{\mnras}{Mon. Not. R. Astron. Soc.}
\newcommand{\aap}{Astronomy \& Astrophysics\xspace}
\newcommand{\apjl}{Astrophysical Journal Letters\xspace}

\def\be{\begin{equation}}\def\ee{\end{equation}}
\def\ba{\begin{eqnarray}}\def\ea{\end{eqnarray}}

\def\({\left(}\def\){\right)}
\def\[{\left[}\def\]{\right]}

\begin{document}

\title{Probing self-interacting dark matter via gravitational-wave background from eccentric supermassive black hole mergers}

\author{Mu-Chun Chen}
\email{chenmuchun23@mails.ucas.ac.cn}

\affiliation{School of Astronomy and Space Science, University of Chinese Academy of Sciences (UCAS), Beijing 100049, China}

\author{Yong Tang}
\email{tangy@ucas.ac.cn}
\affiliation{School of Astronomy and Space Science, University of Chinese Academy of Sciences (UCAS), Beijing 100049, China}
\affiliation{School of Fundamental Physics and Mathematical Sciences, Hangzhou Institute for Advanced Study, UCAS, Hangzhou 310024, China}
\affiliation{International Centre for Theoretical Physics Asia-Pacific, UCAS, Beijing 100190, China}
\begin{abstract}
The nature of dark matter is still mysterious despite various astronomical evidence. As a possible candidate, self-interacting dark matter (SIDM) can potentially resolve some issues appearing in the cold dark matter paradigm. Here we investigate how SIDM around supermassive black holes (SMBHs) in galaxy centers may form a density spike and imprint in the spectrum shape of stochastic gravitational-wave background from SMBH binaries (SMBHBs). Employing a refined dynamical friction formula and consistently evolving the orbital dynamics, we demonstrate that current pulsar timing array (PTA) data are sensitive to the cross section of SIDM with $\sigma(v)/m_\chi\lesssim0.66\,\mathrm{cm}^2/\mathrm{g}$, comparable to other astrophysical probes. We also highlight the importance of including the eccentricity of SMBHBs in the parameter inference, which would affect the results significantly. Our findings reveal the promising potential of PTA observations in probing the nature of dark matter. 

\end{abstract}

\maketitle

\section{Introduction}
\label{sec:intro}
Dark matter is one of the main components of the Universe, with its existence firmly established through multiple astronomical observations including galaxy rotation curves~\cite{1970ApJ...159..379R, Begeman:1991iy}, Bullet Cluster~\cite{2009GReGr..41..207Z, Clowe_2006}, cosmic microwave background anisotropies~\cite{2020}, and large-scale structure~\cite{Dietrich_2012}. Although at the galactic scale and beyond these observations are consistent with the cold dark matter (CDM) paradigm, at small scales or near galactic centers, the particle nature of dark matter remains uncertain. Among numerous dark matter candidates, self-interacting dark matter (SIDM) is a popular and attractive model that may behave differently at small scales~\cite{Dave_2001, Spergel_2000, PhysRevLett.85.1158, Hui_2017}. While no definitive signatures have been detected in terrestrial experiments, any new proposals to probe the nature of dark matter would be warranted. 

Since the first detection of gravitational waves (GWs)~\cite{Abbott_2016}, the advent of gravitational-wave astronomy has opened new detection windows for dark matter. Studies of GW signals with both ground‑based and space‑based detectors—such as LVK~\cite{Martynov_2016, 2015CQGra..32b4001A, akutsu2020overviewkagradetectordesign}, LISA~\cite{amaroseoane2017laserinterferometerspaceantenna}, and Taiji~\cite{Hu:2017mde}—enable discrimination among a variety of dark‑matter candidates, such as CDM~\cite{Blumenthal:1984bp, 1982ApJ...263L...1P}, SIDM~\cite{Kaplinghat_2016, Tulin_2018, adhikari2022astrophysicaltestsdarkmatter, Kong:2025irr, nadler2025sidmconcertocompilationdata, koo2025dynamicalfrictioncircularorbits, banik2025echoesselfinteractingdarkmatter, sarkar2025exploringultralightdarkmatter, xie2025selfinteractioneffectskerrblack, hou2025universalanalyticmodelgravitational, li2025secludedselfinteractingdarksector, Aurrekoetxea_2024, Xie:2025udx}, and ultralight dark matter (ULDM)~\cite{Yu:2023iog, Yao:2024fie, Liu:2021zlt, Wu:2024sgk, Blas:2024duy, Yu:2024enm, Kim:2023kyy}. GW detection has demonstrated remarkable potential for fundamental physics~\cite{Andr_s_Carcasona_2024, Ng_2021, PhysRevD.111.055031, PhysRevD.105.063030, Chen:2024ery, sarkar2025exploringultralightdarkmatter}. Notably, the search for dark matter is expected to become increasingly intertwined with gravitational-wave astronomy.

A promising avenue for gravitational-wave tests of dark matter involves studying binary systems influenced by dark matter surrounding supermassive black holes (SMBHs). In these extreme gravitational environments, dark matter accumulates to form high-density spikes~\cite{Gondolo_1999}, which can significantly modify the inspiral dynamics of black hole binaries. The observational data from pulsar timing arrays (PTAs) such as NANOGrav, EPTA, and PPTA provide compelling evidence for the existence of a stochastic gravitational-wave background (SGWB)~\cite{Aggarwal_2019, Goncharov_2021, Chen_2021}. This is supported by the Hellings-Downs correlation in pulsar timing residuals, indicating sensitivity to nanohertz-frequency GWs, primarily from supermassive black hole binaries (SMBHBs) and potentially exotic sources like cosmic string collisions or neutron star mergers~\cite{Chang_2022, Abbott_2018}.

While the canonical picture predicts that the characteristic strain $h_c$ should scale as $f^{-2/3}$ for a population of inspiraling SMBHBs~\cite{Sesana_2008, Burke_Spolaor_2019}, environmental effects around black holes may alter the merger process, causing deviations of the stochastic gravitational-wave background from the ideal power-law spectrum. The NANOGrav study suggest that environmental effects may play a significant role in supermassive black hole mergers~\cite{Agazie_2023}. For instance, surrounding matter components such as gas, stars, and dark matter can accelerate the energy dissipation of binary systems. More studies have investigated the effects of three-body interactions~\cite{chen2024galaxytomographygravitationalwave}, CDM~\cite{ghoshal2023probingdarkmatterdensity, Hu_2025}, ULDM~\cite{Cai:2024thd, Sun:2021yra, li2025probingbosoncloudssupermassive, Wu_2023} and eccentric orbits on the stochastic gravitational-wave background at low frequencies~\cite{Huerta_2015}.

Notably, SIDM may help to resolve the core-cusp problem of dark matter density spikes~\cite{PhysRevLett.84.3760, Dave_2001, Elbert_2015, Tulin_2018}. Moreover, SIDM models~\cite{alon2024} may also simultaneously address the ``final parsec problem"~\cite{Milosavljevic2003} and contribute additional energy dissipation in binary systems. However, Ref.~\cite{alon2024} has only investigated the simple dynamical friction of SIDM in circular-orbit binary SMBH mergers, which should be eccentric in general scenarios . 
%we examine merger processes with initial eccentricity and more realistic calculations of dynamical friction.

In this work, we adopt the SIDM framework with a velocity‑dependent scattering cross section and incorporate a refined dynamical friction formulation that captures additional drag from high‑velocity dark matter particles. We then investigate how dynamical friction affects the eccentric orbits of SMBHBs, which not only augments gravitational radiation but also induces a characteristic suppression in the SGWB’s spectrum at low frequencies relative to circular binaries. Our analysis furnishes a more realistic physical description. Through Bayesian inference applied to observed characteristic‑strain data, we obtain the parameter fit range of the SIDM model and quantify how initial orbital eccentricity reshapes the posterior distributions of model parameters.

This paper is organized as follows. Section.~\ref{sec.2} presents the methodology for constructing dark matter density and velocity profiles around binary SMBHs, and we compare density profiles under different parameter configurations. Section.~\ref{sec.3} details numerical computations of binary SMBH evolution incorporating eccentric orbits and dynamical friction, with particular emphasis on quantifying contributions from additional drag terms. Section.~\ref{sec.4} demonstrates how variations in SIDM parameters and initial orbital eccentricity affect the characteristic strain spectrum of stochastic gravitational waves, comparing these predictions with PTA observations. In Section.~\ref{sec.5} we employ Bayesian inference and $\chi^2$ analysis to quantify parameter compatibility with observational data, ultimately showcasing the sensitive parameter space of SIDM models.  

\section{SIDM Profile AROUND BLACK HOLES}\label{sec.2}
We aim to establish a suitable dark matter halo model for galaxies of different masses and redshifts, which will be used to calculate the orbital evolution under dynamical friction in the context of supermassive black hole mergers. For SIDM models, the key lies in the construction of an isothermal core. 

%Therefore, we first need to define the density distribution of the outer halo to establish the boundary conditions.

\subsection{OUTER HALO DENSITY}

We consider spherically symmetric dark matter halos and use the Navarro-Frenk-White (NFW) profile~\cite{Navarro_1996} to characterize dark matter distribution in the outer region as a function of radial distance  $r$ from the Galaxy Center,
\begin{equation}\label{eq:NFW}
    \rho_{\mathrm{NFW}}(r)=\frac{\rho_s}{\left( \frac{r}{r_s} \right)\left(1+\frac{r}{r_s}\right)^2},
\end{equation}
where $\rho_s$ and $r_s$ represent the characteristic density and scale radius of the density profile respectively. 

To construct the isothermal core of SIDM around a binary SMBH system with given redshift $z$ and total mass $M_{\mathrm{tot}}=m_1+m_2$, where $m_1$ and $m_2$ are the component masses with mass ratio $q\equiv m_2/m_1$, one must first determine the outer dark matter density profile, specified by $\rho_s$ and $r_s$. Next we shall obtain $\rho_s$ and $r_s$ through the following procedure,
\begin{equation}
    M_{\mathrm{tot}}\to M_{\mathrm{bulge}}\to M_{\star}\to M_{200}\to \rho_s,\,r_s,
\end{equation}
where $M_{\mathrm{bulge}}$, $M_{\star}$, and $M_{200}$ represent the bulge mass, stellar mass, and halo mass of the host galaxy respectively. The $M_{\mathrm{tot}}$--$M_{\rm bulge}$ relation reflects the coevolution of supermassive black holes and their host galaxies~\cite{Kormendy_2013},
\begin{equation}\label{eq:Mtot_Mbulge}
        \mathrm{log}_{10}\left(\frac{M_{\mathrm{tot}}}{\,M_{\odot}}\right)=8.7+1.1\,\mathrm{log}_{10}\left(\frac{M_{\rm bulge}}{10^{11}\,M_{\odot}}\right).
\end{equation}
Furthermore, $M_{\mathrm{bulge}}$ can be expressed as a function of the stellar mass $M_{\star}$ of the host galaxy~\cite{Chen_2019},
\begin{equation}\label{eq:Mbulge_Mstar}
    M_{\rm bulge}=f_{\star,\rm bulge}M_{\star},
\end{equation}
where $f_{\star,\mathrm{bulge}}=0.615 + df_{\star}$, with
\begin{equation}
        df_{\star} = 
        \begin{cases}
        0, & \text{if } M_{\star} \leq 10^{10}\,M_{\odot} \\
        \frac{\sqrt{6.9} \exp \left( \frac{-3.45}{\log_{10} M_{\star} - 10} \right)}{(\log_{10} M_{\star} - 10)^{1.5}}, & \text{if } M_{\star} > 10^{10}\,M_{\odot}.
        \end{cases}
\end{equation}
By combining Eq.~\eqref{eq:Mtot_Mbulge} and Eq.~\eqref{eq:Mbulge_Mstar}, we can numerically solve for $M_{\star}$ from $M_{\mathrm{tot}}$. Then, using the relation between $M_{\star}$ and $M_{200}$, we can determine the mass of the dark matter halo~\cite{Girelli_2020},
\begin{equation}\label{poly1}
        \frac{M_{200}}{M_{\star}}(z) = \frac{1}{2A(z)} \left[ \left( \frac{M_{A}(z)}{M_{200}} \right)^{\beta(z)}+\left( \frac{M_{200}}{M_{A}(z)} \right)^{\gamma(z)} \right],
\end{equation}
where $A(z), M_A(z), \beta(z), \gamma(z)$ are polynomials with respect to redshift $z$, from which we derive the virial mass. We define $R_{200}$ as the radius within which the enclosed mass equals the virial mass under the NFW density profile, where this mass is equivalent to 200 times the critical density $\rho_c(z)$ multiplied by the volume enclosed within $R_{200}$~\cite{Navarro_1997},
\begin{equation}\label{mass200}
    M_{200}=\frac{4\pi {{R}^3}_{200}}{3}\cdot 200\cdot\rho_c(z)=4\pi\int\mathrm{d}r\, r^2\rho_{\mathrm{NFW}}(r).
\end{equation}
The scale radius $r_s$ of the NFW density profile can be determined through the relationship between the concentration parameter $c_{200}$ and the virial radius $r_{200}$~\cite{Klypin_2016},
\begin{equation}\label{eq:poly2}
    c_{200}(z) = \frac{R_{200}}{r_s}=\frac{C_{c}(z)}{\left(M_{200}/M_{\mathrm{ref}}\right)^{\gamma_{c}(z)}} \left[ 1 + \left( \frac{M_{200}}{M_{c}(z)} \right)^{0.4} \right],
\end{equation}
where $M_{\mathrm{ref}}$ is defined as $10^{12}\,h^{-1}M_{\odot}$, and the remaining parameters $C_c, \gamma_c, M_c$ are expressed as functions of redshift. We have compiled the polynomials mentioned here, along with those referenced in Eq.~\eqref{poly1}, in Appendix.~\ref{App:mass_ploynomials}. Once we obtain $r_s$, we can inversely solve for $\rho_s$ through Eq.~\eqref{mass200}.

\subsection{DENSITY AND VELOCITY DISTRIBUTION OF SIDM}\label{cap:profile_theory}

We consider dark matter models with velocity-dependent cross section. The self-interacting cross section of dark matter can be written as~\cite{alon2024}
\begin{equation}
    \sigma(v)=\frac{\sigma_0}{1+\left(\frac{v}{v_t}\right)^4}.
\end{equation}
Here $v_t$ denotes the transition velocity, defined as the mass ratio between mediator and dark matter $v_t \equiv m_{A'}/m_{\chi}$ and $v$ denotes the velocity dispersion of the dark matter particles. Here we do not specify the particle physics model of dark matter~\cite{Holdom:1985ag, Pospelov_2009, essig2013darksectorsnewlight, Buckley_2010, Tulin_2013, Tulin_2013.2, Tulin_2018, Tulin_2013, Cline_2014, Boddy_2016, Feng_2010, Agrawal_2017, PhysRevLett.90.225002, PhysRevE.70.056405, Ko_2014, Choi_2017, Wang:2022lxn, yang2025diversifyinghalostructurestwocomponent, Boehm:2020wbt, Bernal:2019uqr}, as the relevant physical parameters are represented by $\sigma_0$ and $v_t$.
%The detailed derivation process can be found in Appendix.~\ref{App:self-interaction_cross section}. 
It can be seen that the cross section behaves like a constant at low velocities, while at high velocities it transitions to a behavior similar to Coulomb scattering, $\sigma(v) \propto 1/v^4$.

Owing to the large cross section of SIDM, the exchange of momentum and energy among particles within a SIDM halo occurs rapidly. In the region with radial coordinate $r$ smaller than some value $r_1$, the dark matter density tends to flatten, forming an isothermal core. The boundary $r_1$ of the isothermal core is defined through the average interaction cross section~\cite{Kaplinghat_2016, alon2024},
\begin{equation}\label{r1_solver}
    \frac{\langle\sigma(v)v\rangle}{m_{\chi}}\cdot\rho_{\mathrm{NFW}}(r_1)\cdot t_{\mathrm{age}}\approx 1,
\end{equation}
where $t_{\mathrm{age}}$ represents the age of the galaxy core. This indicates that dark matter particles within the radius $r_1$ have at least one interaction during the galaxy merger process. We use $t_{\mathrm{age}}=1\,\mathrm{Gyr}$ in this paper for convenience. This does not affect our final analysis of the parameter $\sigma_0/m_{\chi}$, as we can easily rescale to different $t_{\mathrm{age}}$ by adjusting the size of $\sigma_0/m_{\chi}$ and compute the cross section for different characteristic velocities.

In the region where $r < r_1$, we consider the velocity distribution of dark matter particles is isotropic~\cite{Kaplinghat_2016}, and the velocity dispersion within the isothermal core is approximately constant which has also been corroborated by N-body simulations~\cite{Rocha_2013, Elbert_2015, Vogelsberger_2012}. By incorporating the equation of state $P=\rho v^2_0$ for fluids and the hydrostatic equilibrium condition $\mathbf{\nabla}P=-\rho\mathbf{\nabla}\Phi$ into the Poisson equation~\cite{Kaplinghat_2016, Tulin_2018}, we can obtain
\begin{equation}\label{findv0}
    v^2_0{\nabla}^2\mathrm{ln}\rho=-4\pi G\rho,
\end{equation}
where $v_0$ represents the velocity dispersion in the isothermal core. 

Using ${\rho}' (0)=0$, $\rho(r_1)=\rho_{\mathrm{NFW}}(r_1)$ as boundary conditions, and requiring that the mass enclosed within the isothermal core matches the result from the NFW density profile
\begin{equation}
    \int_{0}^{r_1}\frac{r^2\rho_s}{\left(\frac{r}{r_s}\right)\left(1+\frac{r}{r_s}\right)}\mathrm{d}r=\int_0^{r_1}r^2\rho(r)\mathrm{d}r,
\end{equation}
we can obtain the density distribution within the isothermal core.

\begin{figure}[t]
    \centering
    \includegraphics[width=0.45\textwidth]{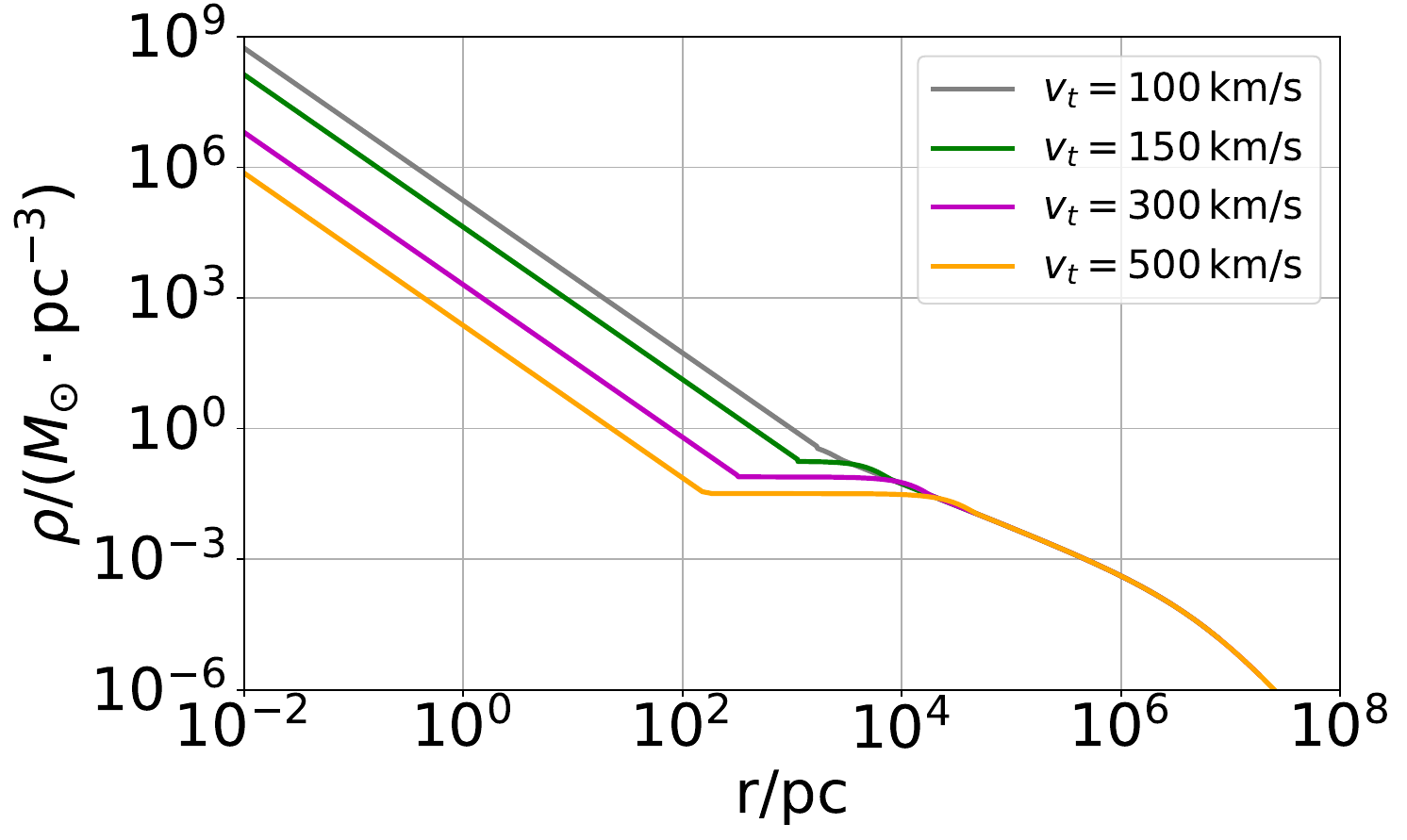} 
    \vspace{15pt} 
    \includegraphics[width=0.45\textwidth]{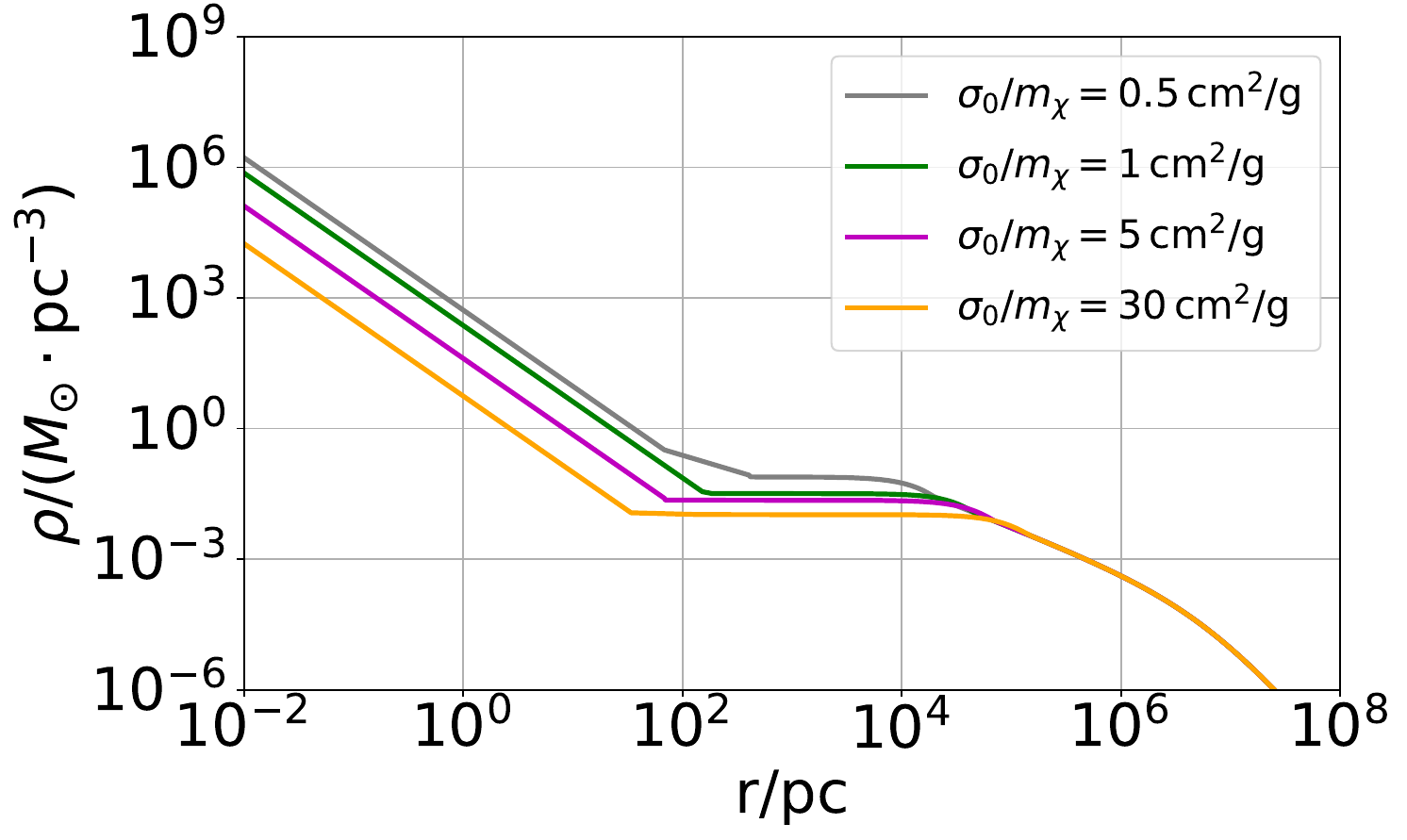}
    \vspace{-15pt}
    \caption{We take the parameters of the binary supermassive black hole system as the example: $M_{\mathrm{tot}}=10^{10}\,M_{\odot}$, $z=0$, $q=0.5$. Upper: the dark matter density profile is plotted for different transition velocities while fixing the dark matter self-interacting cross section $\sigma_0/m_{\chi} = 1\, \mathrm{cm}^2/\mathrm{g}$. Lower: the dark matter density profile with a different cross section while fixing the transition velocity $v_t = 500\, \mathrm{km}/\mathrm{s}$. Here, the region $r < 10^2\, \mathrm{pc}$ corresponds to the spike, the flat middle region is the isothermal core, and the $r > 10^6\, \mathrm{pc}$ segment belongs to the NFW region.}
    \label{fig:differ_rho_vt_sigma0}
\end{figure}

By solving Eq.~\eqref{r1_solver} and Eq.~\eqref{findv0}, we have determined the velocity dispersion $v_0$ of dark matter particles and the radius of the isothermal core $r_1$ within the isothermal region. Then we define the black hole influence radius as~\cite{Gondolo_1999}
\begin{equation}
    r_{\mathrm{sp}}=\frac{GM_{\mathrm{tot}}}{v^2_0}.
\end{equation}
This also represents the radius of the dark matter spike. For CDM, the density distribution of the dark matter spike can be described by a power law~\cite{Ullio_2001, Merritt_2002, Gondolo_1999}. In the case of SIDM, under the nonrelativistic approximation, the density and velocity distribution can also be expressed in the form of a power law~\cite{Shapiro_2014, Sadeghian_2013, alon2024},
\begin{align}
    \rho_{\mathrm{sp}}(r)&=\rho_0\left(\frac{r_{\mathrm{sp}}}{r}\right)^{\frac{3+a}{4}},\\
    \frac{v(r)}{v_0}&=\frac{7}{11}+\frac{4}{11}\left(\frac{r_{\mathrm{sp}}}{r}\right)^{\frac{1}{2}}.
\end{align}
The exponent $a$ represents the dependence of the interaction cross section on the velocity power. For the dark photon model, the interaction cross section approximates to a constant $(a=0)$ at low velocities, while at high velocities, $a=4$.

We note that for certain masses of central black holes and redshifts, only a sufficiently large interaction cross section $\sigma_0/m_{\chi}$ can ensure that the radius of the dark matter spike $r_{\mathrm{sp}}$ exceeds the radius $r_1$ of the isothermal core. For dark matter spikes that have lost their isothermal core, the stable structure may be disrupted due to perturbations caused by the loss of orbital energy during binary inspiral. Meanwhile, a very large cross section may also lead to core collapse or a weakening of dynamical friction effects, causing the binary merger timescale to exceed the Hubble time~\cite{alon2024, Zhang:2024fib, Elbert_2015}.

We define the transition radius $r_t$ such that $v(r_t)=v_t$; the density distribution within the spike is given by
\begin{equation}
    \rho_{\mathrm{sp}} = 
    \begin{cases}
    \rho_0\left(\frac{r_{\mathrm{sp}}}{r_t}\right)^{\frac{3}{4}}\left(\frac{r_t}{r}\right)^{\frac{7}{4}}, & \text{if } r < r_t \\
    \rho_0\left(\frac{r_{\mathrm{sp}}}{r}\right)^{\frac{3}{4}} & \text{if } r > r_t.
    \end{cases}
\end{equation}

In Fig.~\ref{fig:differ_rho_vt_sigma0}, we construct the complete density profile and examine how the density profile responds to varying transition velocities $v_t$. A distinct flattened density region, corresponding to the isothermal core, appears between approximately $10^2$ and $10^4\,\mathrm{pc}$. For $v_t = 300\,\mathrm{km}\,\mathrm{s}^{-1}$, the one–dimensional velocity dispersion is $v_0 \simeq 373\,\mathrm{km}\,\mathrm{s}^{-1}$, so $v_0 > v_t$. The corresponding transition radius $r_t$ lies outside the spike radius $r_{\mathrm{sp}}$, where the density slope is $\gamma = 7/4$. Beyond $\sim 10^{6}\,\mathrm{pc}$, the profile reverts to the NFW form. In the lower panel of Fig.~\ref{fig:differ_rho_vt_sigma0}, with $v_t=500\,\mathrm{km}\,\mathrm{s}^{-1}$ fixed, we can see that increasing the cross section would broaden the flat core, which boosts particle velocities and thus lowers the density. For the same fractional variations, $v_t$ affects the density more strongly than $\sigma_0/m_{\chi}$: an order-of-magnitude increase in $v_t$ changes the density by roughly 3 orders of magnitude, whereas an order-of-magnitude change in $\sigma_0/m_{\chi}$ alters the density by only about the same order of magnitude.

To verify the consistency between our model and observational results, we also conduct a benchmark analysis using the Milky Way system. By cross validating our theoretical predictions with observational constraints derived from monitoring the $S2$ stellar orbit around the Galactic Center supermassive black hole $\mathrm{Sgr} \mathrm{A}^*$, we find that the enclosed mass within the $S2$ orbital pericenter, even with the parameters $v_t=100\,\mathrm{km}/\mathrm{s},\,\sigma_0/m_{\chi}=0.3\,\mathrm{cm}^2/\mathrm{g}$ that yield the highest spike density, the density at $r=0.01\,\mathrm{pc}$ is approximately $10^7\,M_{\odot}/\mathrm{pc}^3$, and the enclosed mass does not exceed $100\,M_{\odot}$, 1 order of magnitude smaller than the empirically derived upper limit of $1200\,M_{\odot}$~\cite{2024}. 

\section{DYNAMICAL FRICTION AND ORBITAL EVOLUTION}\label{sec.3}
\label{sec:single}
In CDM spike models, particle velocities are generally much smaller than the orbital velocity around the black hole, and the phase-space probability distribution of dark matter particles decreases with increasing velocity. Therefore, particles occupying velocities between the orbital velocity and the escape velocity constitute a very small fraction, making the simplified Chandrasekhar dynamical friction formula—which neglects the contribution of particles in this velocity range—sufficiently accurate for describing the dynamical friction process~\cite{1943ApJ....97..255C}.

However, for SIDM, the fraction of dark matter particles with high velocities is non-negligible, as we will show below. Thus, it is typically necessary to consider additional frictional corrections arising from particles with velocities between the orbital and escape velocities. Our subsequent calculations indicate that this corrective effect cannot be neglected, particularly for the evolution of orbits with moderate initial eccentricity.

The dynamical friction of dark matter with additional correction can be expressed as~\cite{dosopoulou2024dynamicalfrictiondarkmatter}
\begin{equation}\label{eq:friction}
    \mathbf{F}_{\text{df}} = -4\pi G^2 \rho_{\text{sp}} M_{\text{bh}}^2 \frac{\mathbf{v}}{v_{\text{bh}}^3}\left(\log \Lambda N_1 +N_2\right),
\end{equation}
where $M_{\mathrm{bh}}$ represents the black hole mass, $v_{\mathrm{bh}}$ denotes the black hole velocity and $\Lambda$ is the Coulomb logarithm, which we take as $\log\Lambda=3$ in this work~\cite{Gualandris_2008}. $N_1$ and $N_2$ represent the fraction of dark matter particles with velocities less than the black hole's velocity and that with velocities between the black hole's velocity and the escape velocity, respectively.
\begin{align}\label{eq:N1N2}
    N_1 &= \int_0^{v}4\pi v_{\mathrm{DM}}^2 f(v_{\mathrm{DM}})\mathrm{d}v_{\mathrm{DM}},\\
    N_2 &=\int_v^{v_{\mathrm{esp}}}4\pi v_{\mathrm{DM}}^2 f(v_{\mathrm{DM}})\left[\mathrm{ln}\left(\frac{v_{\mathrm{DM}}+v}{v_{\mathrm{DM}}-v}\right)-\frac{2v}{v_{\mathrm{DM}}}\right]\mathrm{d}v_{\mathrm{DM}}.\label{eq:correct_df_term}
\end{align}
For cold dark matter, one usually takes $N_1=1$ and $N_2=0$. However, for SIDM this ratio varies depending on the position within the binary system.

\begin{figure}[t]
    \centering
    \includegraphics[width=0.96\linewidth]{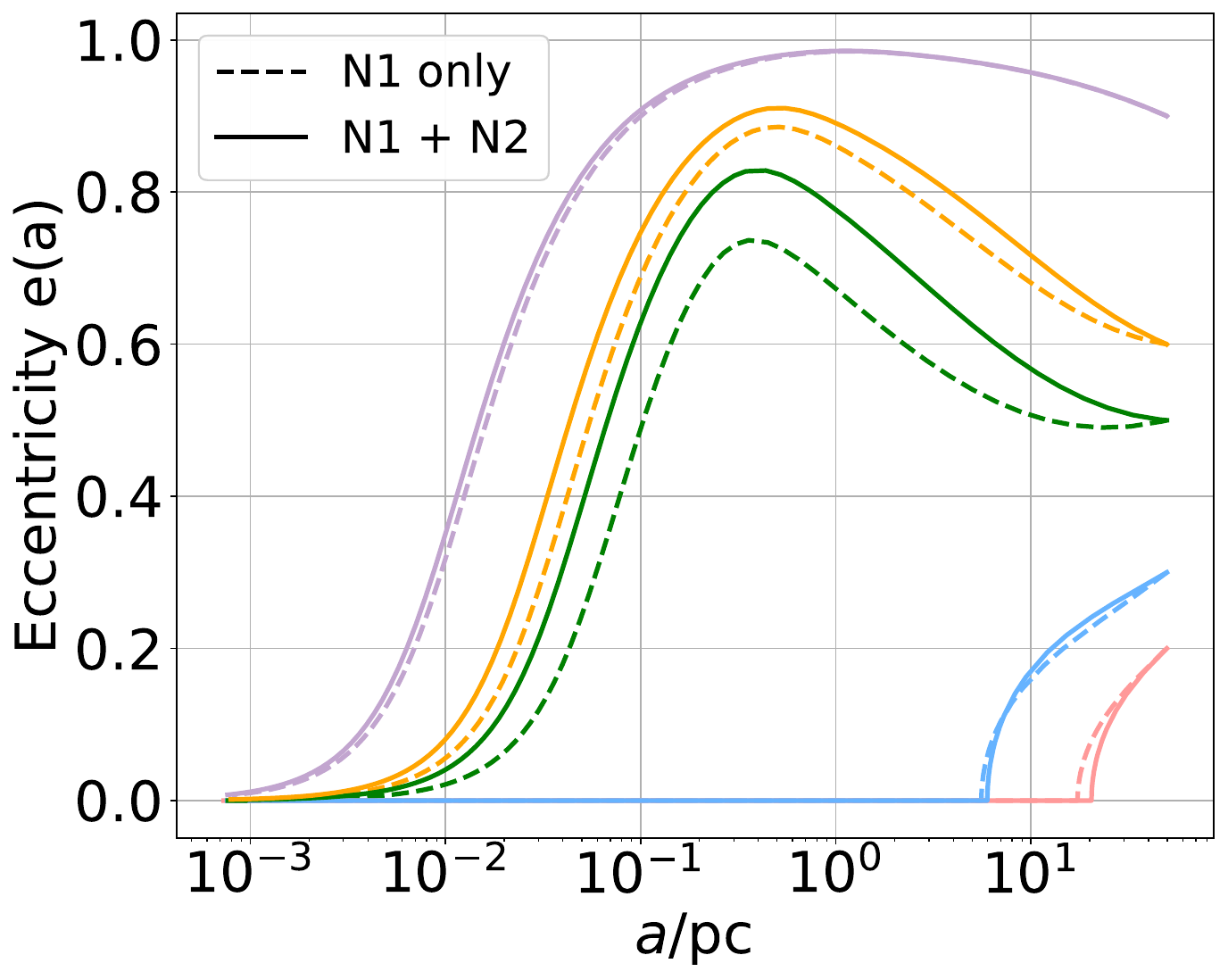}
    \caption{This figure shows the $e(t)$ versus $a(t)$ during the dynamical friction phase of binary supermassive black holes with parameters $M_{\mathrm{tot}}=10^{10}\,M_{\odot}$, $q=0.6$, $z=2$, $v_t=300\,\mathrm{km}/\mathrm{s}$, and $\sigma_0/m_{\chi}=6\,\mathrm{cm}^2/\mathrm{g}$. The dashed curves represent the evolution according to the classical Chandrasekhar formula ($N_2 = 0$), while the solid curves have included the correction term $N_2$. The differently colored curves in the plot correspond to different initial eccentricities.}
    \label{fig:e(a)_e0_3_500}
\end{figure}

Treating all orbits as circular would undoubtedly neglect the modifications in gravitational radiation induced by eccentric orbits, thereby leading to an overestimation of the contribution from dynamical friction. In this paper, we consider orbital evolution with nonzero eccentricity, as opposed to the simplified circular orbit scenario.

We now examine the evolution of the binary system's eccentric orbit under the combined influence of dynamical friction and gravitational radiation. The change rates of the binary system's energy and angular momentum can be expressed as
\begin{align}
    \dot{E} &= \left\langle \frac{\mathrm{d}E}{dt} \right\rangle_{\text{GW}} + \left\langle \frac{\mathrm{d}E}{dt} \right\rangle_{\text{DF}},\\
    \dot{L} &= \left\langle \frac{\mathrm{d}L}{dt} \right\rangle_{\text{GW}} + \left\langle \frac{\mathrm{d}L}{dt} \right\rangle_{\text{DF}},
\end{align}
where the contribution from emitting GWs can be expressed as~\cite{PhysRev.136.B1224}
\begin{align}
    \left\langle \frac{\mathrm{d}E}{dt} \right\rangle_{\text{GW}} &= -\frac{32}{5} \frac{G^4 m_1^2 m_2^2 M}{c^5 a^5 (1 - e^2)^{7/2}} \left( 1 + \frac{73}{24}e^2 + \frac{37}{96}e^4 \right),\\
    \left\langle \frac{\mathrm{d}L}{dt} \right\rangle_{\text{GW}} &= -\frac{32}{5} \frac{G^{7/2} m_1^2 m_2^2 M^{1/2}}{c^5 a^{7/2} (1 - e^2)^2} \left( 1 + \frac{7}{8}e^2 \right).
\end{align}
We use the phase-averaging method to approximate the change rates of thr $i$-th object's energy and angular momentum in the binary system due to dynamical friction,
\begin{align}\label{eq:dE_dL}
    \left\langle \frac{\mathrm{d}E_i}{dt} \right\rangle_{\text{DF}} &= -\frac{1}{T} \int_0^T F_i v_i \, \mathrm{d}t\\
    \left\langle \frac{\mathrm{d}L_i}{dt} \right\rangle_{\text{DF}} &= -\frac{1}{T} \int_0^T F_i r_i \frac{r_i \dot{\phi}_i}{v_i} \, \mathrm{d}t.
\end{align}
The evolution of the orbital semimajor axis and eccentricity as functions of time are determined by the variation of the total energy and angular momentum. Detailed derivations are provided in Appendix~\ref{App:ae_rate}.

We select supermassive black hole merger events within a mass range of $M_{\mathrm{tot}}/\,M_{\odot}\in[10^6, 10^{11}]$, assuming that the inspiral process of the eccentric orbit begins at an average orbital frequency $F_0=10^{-13} \,\mathrm{Hz}$~\cite{hu2025distinctivegwbseccentricinspiraling, fastidio2024eccentricityevolutionptasources}. 
%We posit that when supermassive black hole mergers enter the dark matter dynamical friction phase, they generally retain significant orbital eccentricity. 
High-resolution cosmological $N$-body simulations (IllustrisTNG100-1)~\cite{2016A&A...594A..13P} indicate that approximately $95\%$ of galactic merger orbits exhibit high eccentricities~\cite{fastidio2024eccentricityevolutionptasources, rawlings2023revivingstochasticityuncertaintysmbh}, with binary systems maintaining eccentricities distributed between 0.2 and 0.8 even as they evolve into the PTA frequency band. Therefore, we consider a binary evolution system with different initial eccentricity distribution. Using numerical methods, we can determine the values of the semimajor axis $a(t)$ and eccentricity $e(t)$ at any subsequent time.

\begin{figure*}[t]
    \centering
    \includegraphics[width=0.48\textwidth]{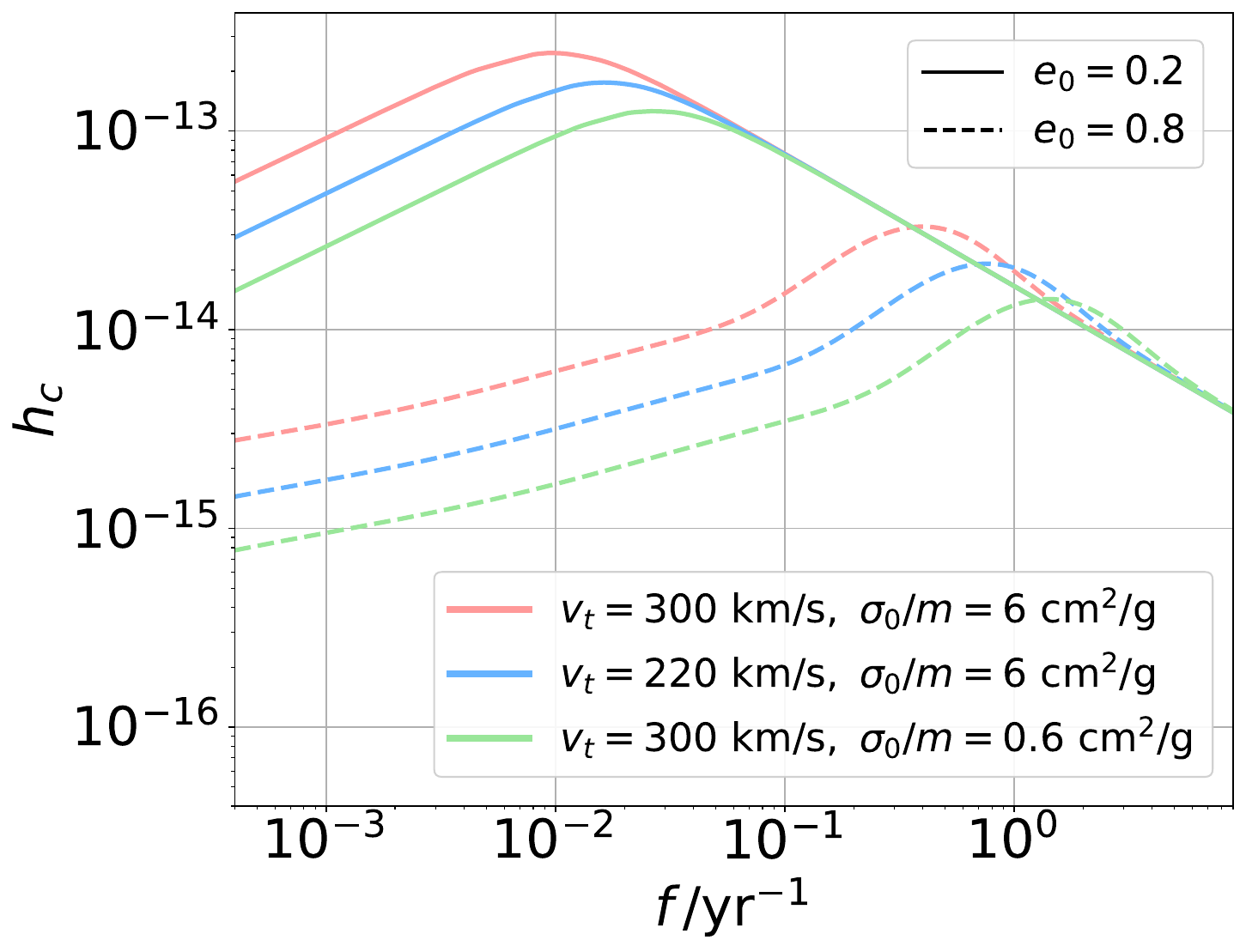}
    \hfill
    \includegraphics[width=0.48\textwidth]{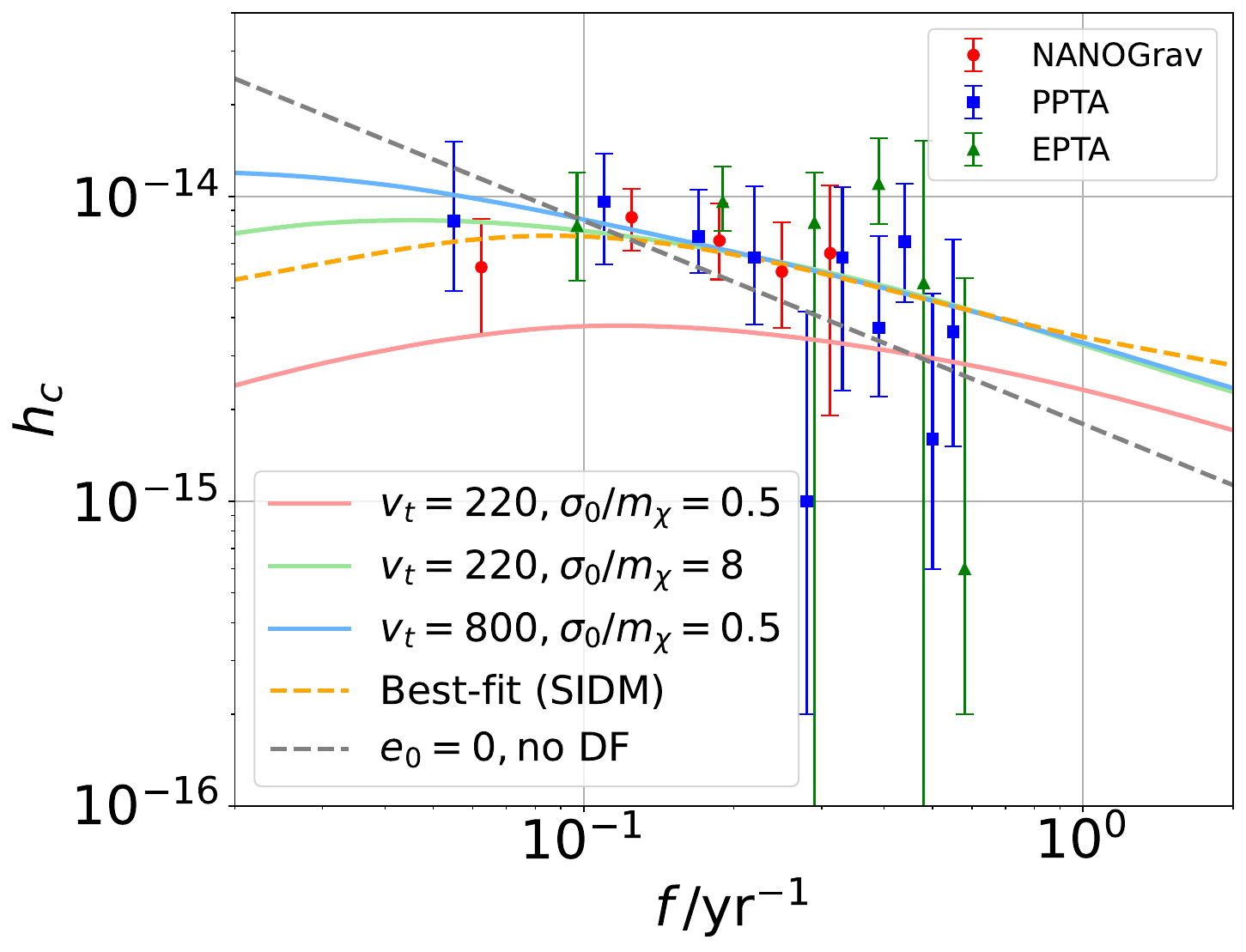} 
    \caption{Left: we present the variation of characteristic strain with dark matter model parameters using a single source as an example. The source parameters are $M_{\mathrm{tot},0}=10^{10} \,M_{\odot}$, $z_0=2$, $q_0=0.6$, with the event rate given by $\mathrm{d}^3N/(\mathrm{d}M_{\mathrm{tot}}\mathrm{d}z\mathrm{d}q)=\delta(M_{\mathrm{tot}}-M_{\mathrm{tot},0})\delta(z-z_0)\delta(q-q_0)$. In the figure, curves of different colors represent characteristic strain with different transition velocities and cross sections, where dashed ones correspond to an initial eccentricity of $0.8$, while the solid ones represent an initial eccentricity of $0.2$. Right: this shows the SGWB characteristic strain  obtained by integrating sources within the comoving volume according to Eq.~\eqref{event_density}, along with the data from PTAs. The solid curves of different colors represent the characteristic strain obtained by averaging over contributions from initial eccentricities for various dark matter model parameters. The orange dashed curve shows the best-fit case between our model and the data, with $v_t = 396.9\,\mathrm{km\,s^{-1}}$, $\sigma_0/m_{\chi} = 2.82\,\mathrm{cm^2\,g^{-1}}$, except for the gray curve, all spectra are normalized with $\psi_0=-2.31$.} The gray dashed line represents the characteristic strain obtained without dynamical friction, assuming purely circular orbits.
    \label{fig:hc_e0}
\end{figure*}
In Fig.~\ref{fig:e(a)_e0_3_500}, we plot the evolution of eccentricity as a function of the orbital semi‐major axis at an initial orbital frequency of $F_0=10^{-13}\,\mathrm{Hz}$, comparing the evolution predicted by the classical Chandrasekhar formula and the modified formula, Eq.~\eqref{eq:friction}, which incorporates additional drag effects from dark matter particles moving faster than the orbiting objects. As we can see, for small initial eccentricities (e.g., $e_0\lesssim0.35$), both dynamical friction and gravitational radiation tend to circularize the orbit. In contrast, for larger initial eccentricities, the orbit is driven by dark matter friction first 
%– including the extra drag effect from self-interacting dark matter particles whose velocities lie between that of the central black hole and the escape velocity – 
toward values close to unity before gravitational radiation dominates and circularizes the orbit.

It is noteworthy that while the evolutionary trajectories predicted by both formulations differ minimally at very high or very low initial eccentricities, the additional $N_2$ term in Eq.~\eqref{eq:correct_df_term} has sizable contribution in driving toward higher eccentricities from intermediate initial values.
This effect significantly impacts the spectrum of GWs during the binary inspiral process, ultimately resulting in a lower characteristic strain within the nanohertz band. 

\section{CHARACTERISTIC STRAIN OF GRAVITATIONAL WAVES}\label{sec.4}
The characteristic strain of the SGWB from supermassive black hole mergers is described by the gravitational wave energy spectrum of the sources and the event rate~\cite{Chen_2017, phinney2001practicaltheoremgravitationalwave},
\begin{equation}
    h^2_c(f)=\frac{4G}{c^2\pi f}\int\mathrm{d}z\mathrm{d}M_{\mathrm{tot}}\mathrm{d}q\frac{\mathrm{d}^3n}{\mathrm{d}z\mathrm{d}M_{\mathrm{tot}}\mathrm{d}q}\frac{\mathrm{d}E_{\mathrm{GW}}}{\mathrm{d}f_s},
\end{equation}
where $f_s=(1+z)f$ takes into account the cosmological redshift of frequency.

The merger rate of supermassive black holes is correlated with the merger rate of galaxies~\cite{Agazie_2023, Chen_2019},
\begin{equation}\label{event_density}
    \frac{\mathrm{d}^3n}{\mathrm{d}z\mathrm{d}M_{\mathrm{tot}}\mathrm{d}q}=\frac{\mathrm{d}^3n_g}{\mathrm{d}z\mathrm{d}M_{\star}\mathrm{d}q_{\star}}\frac{\mathrm{d}M_{\star}}{\mathrm{d}M_{\mathrm{tot}}}\frac{\mathrm{d}q_{\star}}{\mathrm{d}q},
\end{equation}
where $M_{\star}$ is the mass of the more massive galaxy in the initial merger, and $q_{\star}$ is the mass ratio of the two galaxies. This event rate will be described by a series of polynomials,
\begin{equation}\label{event_density_of_galaxy}
    \frac{\mathrm{d}^3n_g}{\mathrm{d}z\mathrm{M_{\star}}\mathrm{d}q_{\star}}=\frac{\Psi(M_{\star},z^{'})}{M_{\star}}\frac{P(M_{\star},q_{\star},z^{'})}{T_{\mathrm{g-g}}(M_{\star},q_{\star},z^{'})}\frac{\mathrm{d}t}{\mathrm{d}z^{'}}.
\end{equation}
The details of the polynomials are elaborated upon in Appendix~\ref{App:number_density}.

For eccentric orbits, the frequency of GWs is no longer simply twice the orbital frequency. For each source, we calculate the contributions from all orbital harmonics~\cite{PhysRev.131.435}:
\begin{equation}
    \frac{\mathrm{d}E_{\text{GW}}}{\mathrm{d}f_s} \bigg|_{f_s = (1+z)f} = \sum_{n=1}^{\infty} \frac{\mathrm{d}E_{\text{GW}}^n/\mathrm{d}t}{n \mathrm{d}f_{\text{orb}}^n/\mathrm{d}t}\bigg|_{f_{\text{orb}}^n = (1+z)f/n} ,
\end{equation}
where
\begin{align}
    \frac{\mathrm{d}E_{\text{GW}}^n}{\mathrm{d}t} &= \frac{32 G^4 M^5}{5 c^5 a^5} \frac{q^2}{(1+q)^4} g(n, e),\\
    \frac{\mathrm{d}f_{\text{orb}}^n}{\mathrm{d}t} &= -\frac{3\sqrt{GM}}{4\pi a^{5/2}} \frac{\mathrm{d}a}{\mathrm{d}t}.
\end{align}
The value of the semimajor axis $a^3=GM_{\mathrm{tot}}/(2\pi f^n_{\mathrm{orb}})$ is given by Kepler's law, and $g(n,e)$ is defined as~\cite{PhysRev.131.435}
\begin{align}
&g(n, e) = \frac{n^4}{32} \bigg[
\big{\{} J_{n-2}(ne) - 2e J_{n-1}(ne) + \frac{2}{n} J_n(ne)  \notag \\
&\;  + 2e J_{n+1}(ne) - J_{n+2}(ne) \big{\}}^2 + \frac{4}{3n^2} J_n^2(ne) \notag \\
&\; + (1 - e^2) \left\{ J_{n-2}(ne) - 2J_n(ne) + J_{n+2}(ne) \right\}^2   \bigg].
\end{align}
For numerical solutions, the value of $n$ cannot reach infinity, but the contribution of harmonics converges as $n$ increases. We define a truncation $n_{\mathrm{max}}$ and ignore the contributions from harmonics when $n$ exceeds $n_{\mathrm{max}}$. Here, $n_{\mathrm{max}}$ is given by~\cite{Hamers_2021}
\begin{equation}
    n_{\mathrm{max}}=3n_{\mathrm{peak}}(e)\approx 6\left(1+\sum_{k=1}^4 c_ke^k_{\mathrm{max}}\right)\left(1+e_{\mathrm{max}}\right)^{-\frac{3}{2}},
\end{equation}
where $c_1=-1.01678$, $c_2=5.57372$, $c_3=-4.9271$, $c_4=1.68506$. We note that $e_{\mathrm{max}}$ is not necessarily the initial eccentricity. Under certain conditions, the eccentricity may increase first, so $e_{\mathrm{max}}$ represents the maximum eccentricity reached during the entire evolution.

In Fig.~\ref{fig:hc_e0}, we present the characteristic strain for individual sources across different initial eccentricities, and we calculate the GW background characteristic strain curves by integrating the event rate. Furthermore, we compare these results with the characteristic strains for pure gravitational radiation and for circular orbits. The left panel demonstrates that higher eccentricity orbits yield lower characteristic strains at frequencies below approximately $1\,\mathrm{yr}^{-1}$, while exhibiting a slight enhancement at higher frequency. This is because higher initial eccentricities delay the onset of gravitational radiation dominance. Additionally, at even higher frequencies, the curves for different eccentricities tend to converge, approaching the formula $f^{-2/3}$ due to gravitational radiation dominance in the region.

In the right panel we consider a uniformly distributed initial eccentricity $e_0=U\left[0.01,0.99\right]$, and we average over all eccentricities to obtain the characteristic strain curves of the SGWB corresponding to different parameters in the right figure. Except for the gray curve that represents the characteristic strain spectrum for purely circular orbits without environmental effects, we fix the normalization parameter $\psi_0 = -2.31$ for all spectra to facilitate a direct comparison of the spectral shapes. It can be observed that due to higher transition velocities $v_t$ and self-interaction $\sigma_0/m_{\chi}$ leading to lower spike densities, the dynamical friction effect is weakened, resulting in higher characteristic strain.

\section{PARAMETER ESTIMATION FOR THE DARK PHOTON MODEL}\label{sec.5}
Based on the pulsar timing residual data from NANOGrav~\cite{Agazie_2023_sgwbsignal}, PPTA~\cite{Reardon_2023}, and EPTA~\cite{antoniadis2024seconddatareleaseeuropean}, we can obtain the central value and uncertainty distribution of the SGWB’s characteristic strain~\cite{10.21468/SciPostPhys.10.2.047, shen2023darkmatterspikesurrounding},
\begin{equation}
    \mathrm{residual}(f)=\frac{1}{4\pi^2 f_{\mathrm{yr}}}\left(\frac{f}{f_{\mathrm{yr}}}\right)^{-\frac{3}{2}}h_c(f),
\end{equation}
where $f$ denotes the characteristic strain frequency of the SGWB, with $f_{\mathrm{yr}} \approx 3.17 \times 10^{-8}\,\mathrm{Hz}$ defining the reference frequency. For our model, the parameters are $\Theta=(v_t, \frac{\sigma_0}{m_{\chi}}, \psi_0)$. Then we can use the maximum likelihood method to construct the likelihood function,
\begin{equation}
    -2 \ln \mathcal{L}(\Theta) = \sum_{\substack{\{\text{EPTA, PPTA}\\ \text{NANOGrav}\}}}
    \left[ \frac{h_{c,i} - h_c(f_i; \Theta)}{\sigma_i} \right]^2.
\end{equation}
Here $h_{c,i}$ is the characteristic strain and $\sigma_i$ is uncertainty, while $h_c(f_i, \Theta)$ represents the theoretically predicted values as a function of the model parameters $\Theta$. Therefore, the posterior distribution of the parameters can be expressed as
\begin{equation}
    P\left(\Theta|h_{c,i}\right)=\frac{P\left(h_{c,i}|\Theta\right)P\left(\Theta\right)}{P\left(h_{c,i}\right)}\propto\mathcal{L}(\Theta)P(\Theta),
\end{equation}
where we set the prior distributions
\begin{align}
    P(\sigma_0/m_{\chi})/{(\mathrm{cm}^2\cdot\mathrm{kg}^{-1})}&=U\left[0.3, 16\right],\\
    P(v_t)/{(\mathrm{km}\cdot\mathrm{s}^{-1})}&=U\left[100,1000\right],\\
    P(\psi_0)&=\mathcal{N}\left[-2.56,0.4\right].
\end{align}
The selection of the parameter prior ranges is based on the self-consistency of this model as well as previous observational or simulation studies. The NANOGrav dataset demonstrates that $\psi_0$ values in the range of $-3$ to $-1.5$ are physically plausible when accounting for astrophysical environmental effects~\cite{Agazie_2023}.

As mentioned in Sec.~\ref{sec.2}, the transition speed $v_t$ is proportional to the mass ratio between the mediator and dark matter. Based on $N$-body simulations of SIDM halos and observational data of dark halos in dwarf galaxies, galaxies, and galaxy clusters, it is shown that the mass ratio is typically on the order of $10^{-3}$~\cite{Kaplinghat_2016, Kahlhoefer_2015}. Additionally, some studies suggest that for SIDM, assuming a constant cross section, the numerical limit typically ranges from $0.1-10\,\mathrm{cm}^2/\mathrm{g}$~\cite{Tulin_2018, Elbert_2018, Kahlhoefer_2015, Jee_2014, Kahlhoefer_2013, Spethmann_2017}. In the environment of supermassive black hole mergers considered in this paper, where a velocity-dependent cross section is considered, the cross section is generally required to be less than $1-2\,\mathrm{cm}^2/\mathrm{g}$ when the dark matter particle velocity reaches $1000 \,\mathrm{km}/\mathrm{s}$ or even higher~\cite{Peter_2013, Harvey_2015, Wittman_2018}.

An excessively large cross section may result in an overly flattened core, weakening dynamical friction effects. This could cause the binary hardening timescale to exceed $10\,\mathrm{Gyr}$, which would be unable to resolve the final parsec problem. For instance, in a system characterized by parameter $M_{\mathrm{tot}}=10^9\,M_{\odot},\,q=0.1,\,e_0=0.1,\,v_t=1000\,\mathrm{km}/\mathrm{s},\,\sigma_0/m_{\chi}=20\,\mathrm{cm}^2/\mathrm{g}$, the calculated dynamical friction timescale $t_{\mathrm{df}}\approx20\,\mathrm{Gyr}$, which exceeds the Hubble time. Considering the possible transition speed on the order of $10^2-10^3\,\mathrm{km}/\mathrm{s}$ and the additional effect of orbital eccentricity, we set the cross section in the range of $0.3-16\,\mathrm{cm}^2/\mathrm{g}$ as a reasonable prior.

\begin{figure}[t]
    \centering
    \includegraphics[width=0.40\textwidth]{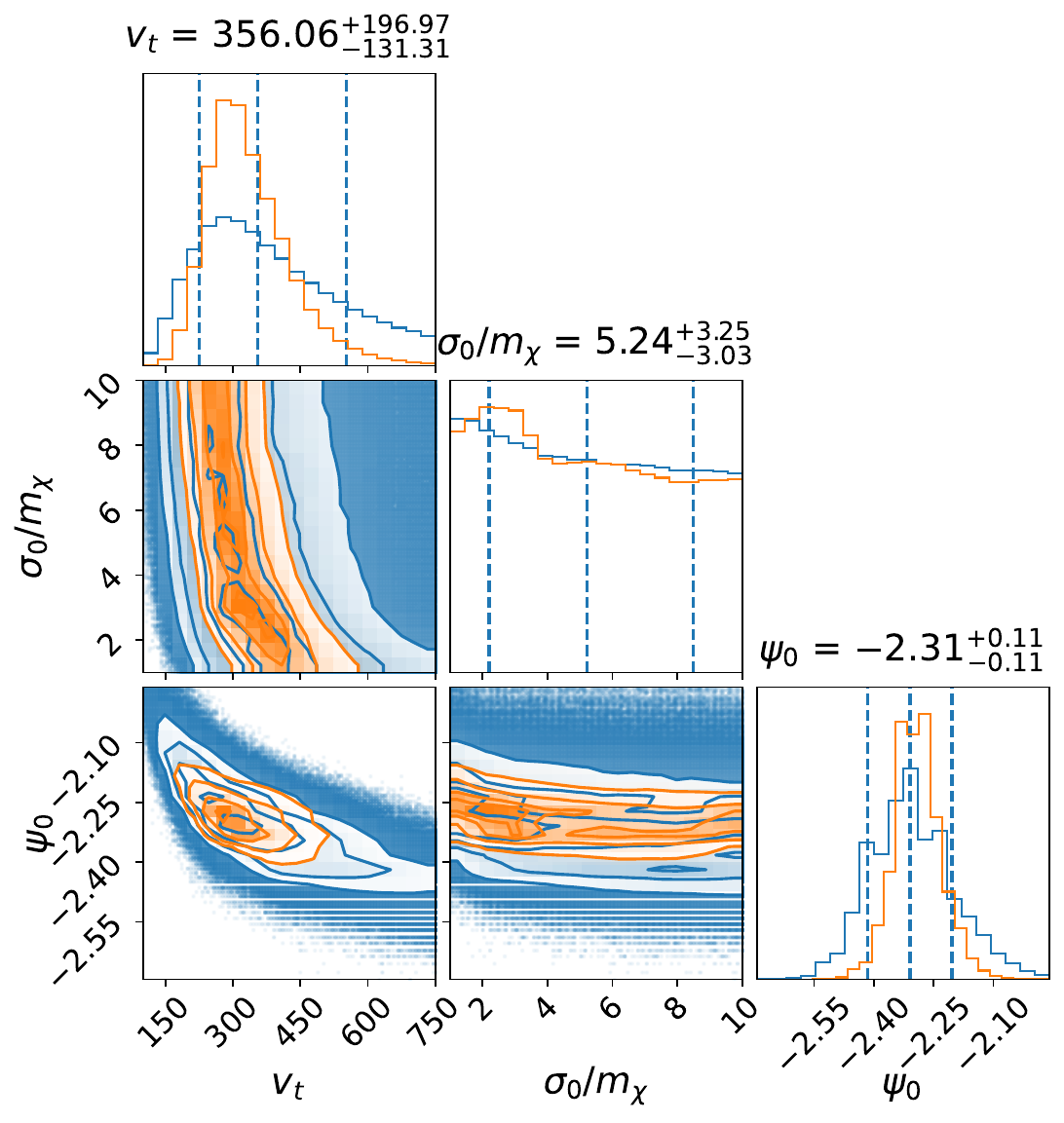} 
    \vspace{-5pt} 
    \includegraphics[width=0.40\textwidth]{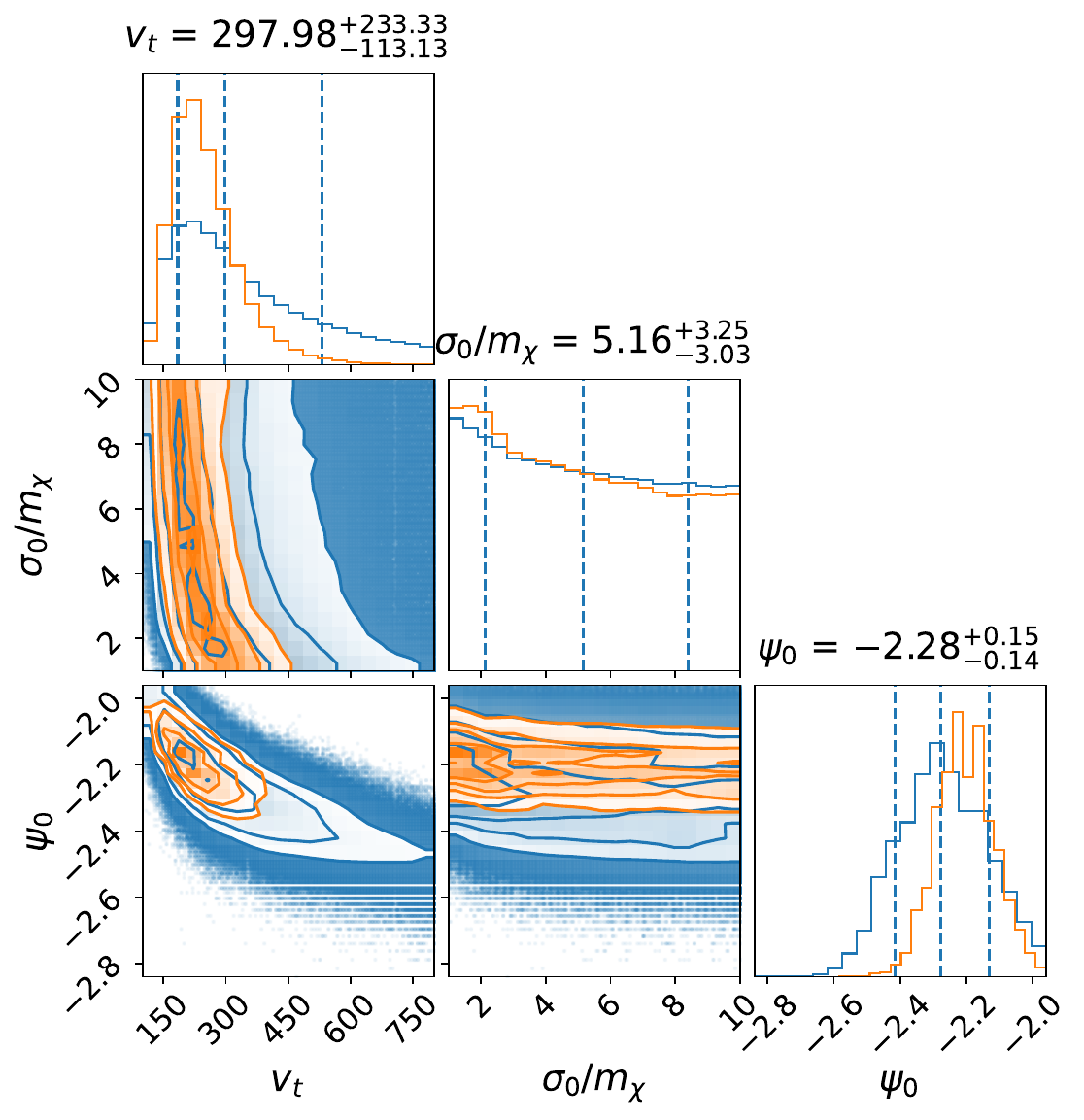}
    \vspace{-8pt}
    \caption{We averaged the characteristic strain over different initial eccentricities. The figure shows the posterior distributions of the parameters $v_t$, $\sigma_0/m_{\chi}$, and $\psi_0$. The upper figure uses the lowest five frequency bins from the NANOGrav 15-year dataset, ten from PPTA, and six from EPTA; in the lower figure, the PPTA and EPTA data remain unchanged, while the NANOGrav data are extended to the lowest 14 frequency bins.}
    \label{fig:result_paras}
\end{figure}
The posterior results of the parameter estimation are presented in Fig.~\ref{fig:result_paras}.
The blue region shows the posterior based on current $h_{c,i}$ uncertainties, while the orange region is the projection when these uncertainties are halved in the future. We integrated the contribution of initial eccentricity with a uniform distribution, averaging the parameter estimation offsets in dark matter models across different initial eccentricities. We have also computed a scenario with a high‐eccentricity normal distribution for initial eccentricity as shown in Fig.~\ref{fig:postdistri_NormalD}.

In our eccentricity-averaged posterior results, the posterior prefers the transition velocity $v_t$ around the $1\sigma$ interval $356.06^{+196.9}_{-131.3}$ ($68\%$ CL), and the normalization constant $\psi_0$ to be $-2.31^{+0.11}_{-0.11}$. Note that the parameter $\sigma_0/m_{\chi}$ does not represent the actual interaction cross section, but requires rescaling to $1\,\mathrm{Gyr}$ for comparison across different core ages, yielding an actual range of $\sigma_0/m_{\chi}\cdot(t_{\mathrm{age}}/1\,\mathrm{Gyr}) \simeq 5.24^{+3.25}_{-3.03}$. Although the overall posterior distribution shows no obvious degeneracy between the cross section and the transition velocity, a trend emerges when the $\sigma_0/m_{\chi}$ falls below $1\,\mathrm{cm}^2/\mathrm{g}$: $\sigma_0/m_{\chi}$ and $v_t$ vary inversely. This occurs because, in this low-$\sigma_0/m_{\chi}$ regime, the impact of $\sigma_0/m_{\chi}$ on $h_c$ becomes comparable to that of the transition velocity. And as the PTA data errors decrease, the inference improves such that a distinct peak emerges in its posterior distribution.

To ensure the reliability of our model, we also calculated the chi-square values between the theoretical predictions and observational data. Using our best-fit parameters $v_t = 396.9\,\mathrm{km\,s^{-1}}$, $\sigma_0/m_{\chi} = 2.82\,\mathrm{cm^2\,g^{-1}}$, and $\psi_0 = -2.31$, and the lowest five frequency bins from the NANOGrav 15-year dataset, ten from PPTA, and six from EPTA, we obtain a minimum chi square $\chi^2_{\mathrm{min}} = 15.588$. Additionally, adopting the parameter set corresponding to the central values from the posterior distributions gives a chi square $\chi^2 = 16.344$. Compared to the scenario without dark matter or when considering eccentric orbital evolution alone, which results in $\chi^2_{\mathrm{GW}} = 21.58$, our best-fit and central posterior scenarios exhibit sizable improvements.

\begin{table*}[t]
    \caption{The table summarizes the posterior estimates of the transition velocity $v_t$ and the cross section normalization $\sigma_0/m_{\chi}$ in the SIDM model, evaluated at a core age of $1\,\mathrm{Gyr}$ for different assumed initial eccentricities. Values marked with the $\diamondsuit$ symbol correspond to the case where the uncertainties in the PTA data are reduced by a factor of 2 in the future. The last two rows present posterior estimates obtained by averaging over contributions from different initial eccentricities and by assuming a normal distribution for the initial eccentricity, respectively.}
    \begin{ruledtabular}
        \renewcommand{\arraystretch}{1.5} % 增加行距
        \begin{tabular}{c|cc|cc}
            $e_0$ & $v_t$ & $\sigma_0/m_{\chi}\cdot(t_{\mathrm{age}}/1\,\mathrm{Gyr})$ & $\diamondsuit\,v_t$ & $\diamondsuit\,\sigma_0/m_{\chi}\cdot(t_{\mathrm{age}}/1\,\mathrm{Gyr})$ \\
            \hline
            $0.1$ & $198.99^{+183.8}_{-63.64}$ & $4.86^{+3.40}_{-2.80}$ & $163.64^{+42.42}_{-28.28}$ & $5.24^{+3.25}_{-3.10}$ \\
            $0.2$ & $198.99^{+183.8}_{-63.64}$ & $4.86^{+3.40}_{-2.80}$ & $163.64^{+42.42}_{-28.28}$ & $5.24^{+3.25}_{-3.10}$ \\
            $0.3$ & $269.70^{+296.9}_{-91.92}$ & $5.54^{+3.10}_{-3.25}$ & $213.13^{+56.57}_{-35.35}$ & $5.24^{+3.18}_{-3.18}$ \\
            $0.4$ & $354.55^{+127.3}_{-77.78}$ & $5.16^{+3.33}_{-3.03}$ & $326.26^{+70.71}_{-42.42}$ & $4.63^{+3.71}_{-2.42}$ \\
            $0.5$ & $425.25^{+127.3}_{-77.78}$ & $5.16^{+3.18}_{-3.10}$ & $404.04^{+77.78}_{-49.49}$ & $5.46^{+2.50}_{-3.03}$ \\
            $0.6$ & $495.96^{+155.6}_{-120.2}$ & $5.16^{+3.25}_{-2.95}$ & $481.82^{+148.5}_{-84.85}$ & $4.78^{+3.63}_{-2.87}$ \\
            $0.7$ & $545.45^{+155.6}_{-141.4}$ & $5.61^{+2.95}_{-3.10}$ & $538.38^{+127.3}_{-98.99}$ & $5.84^{+2.87}_{-3.40}$  \\
            $0.8$ & $566.67^{+148.5}_{-141.4}$ & $5.39^{+3.18}_{-3.10}$ & $616.16^{+134.3}_{-148.5}$ & $3.19^{+5.07}_{-1.66}$ \\
            $0.9$ & $672.73^{+91.92}_{-134.3}$ & $5.24^{+3.25}_{-2.57}$ & $757.58^{+35.35}_{-70.71}$ & $3.04^{+2.12}_{-0.61}$ \\
            \hline
            $U[0.01,0.99]$ & $356.06^{+196.9}_{-131.3}$ & $5.24^{+3.25}_{-3.03}$ & $316.67^{+105.1}_{-65.66}$ & $5.08^{+3.33}_{-2.87}$ \\
            $\mathcal{N}[0.7,0.3]$ & $524.24^{+155.6}_{-134.3}$ & $5.31^{+3.18}_{-3.03}$ & $510.10^{+169.7}_{-98.99}$ & $4.55^{+3.86}_{-2.80}$ \\
        \end{tabular}
    \end{ruledtabular}
    \label{Tab:e0_parameter_distribution}
\end{table*}
\begin{figure}[t]
    \centering
    \includegraphics[width=0.45\textwidth]{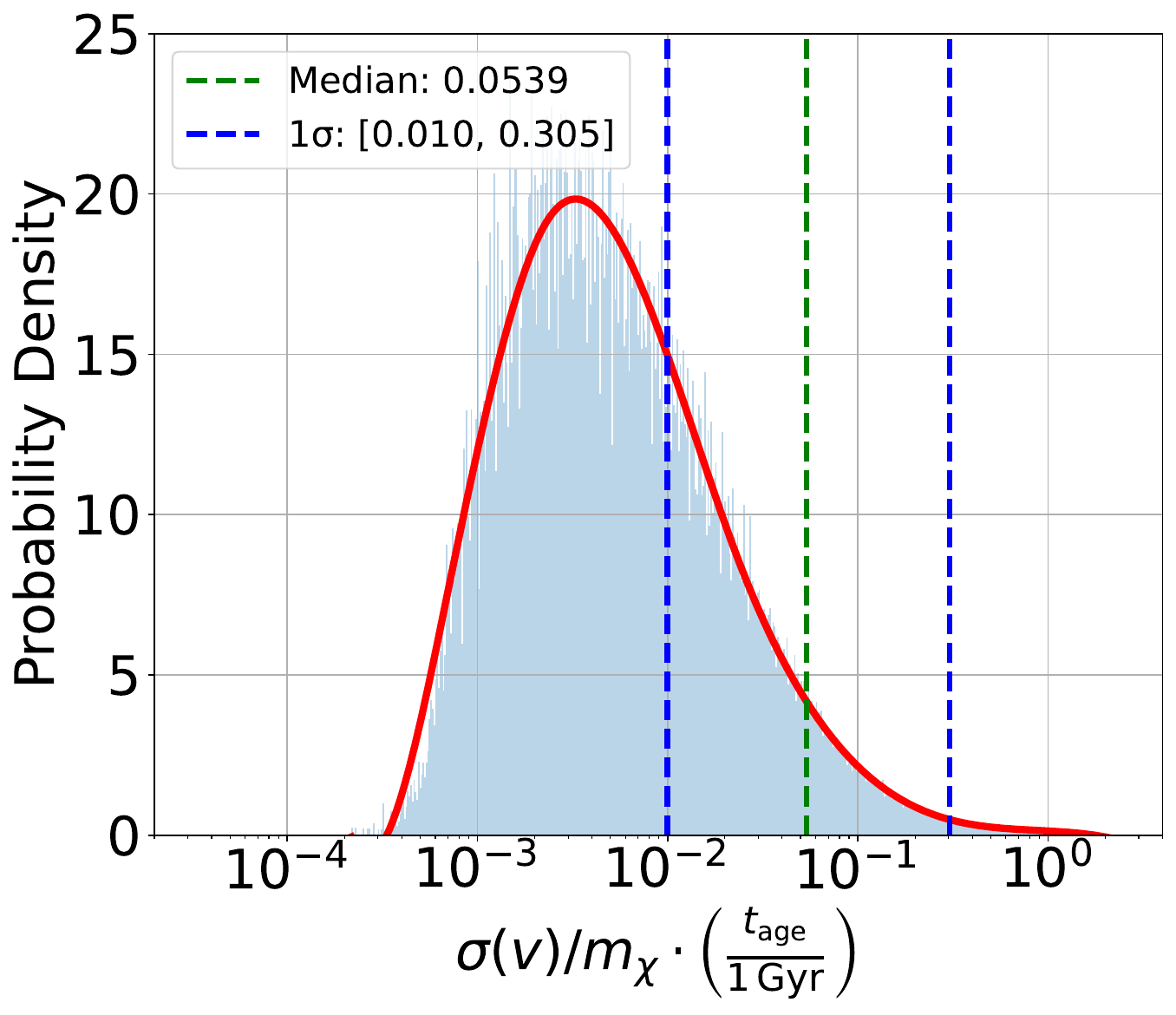} 
    \vspace{-5pt} 
    \includegraphics[width=0.45\textwidth]{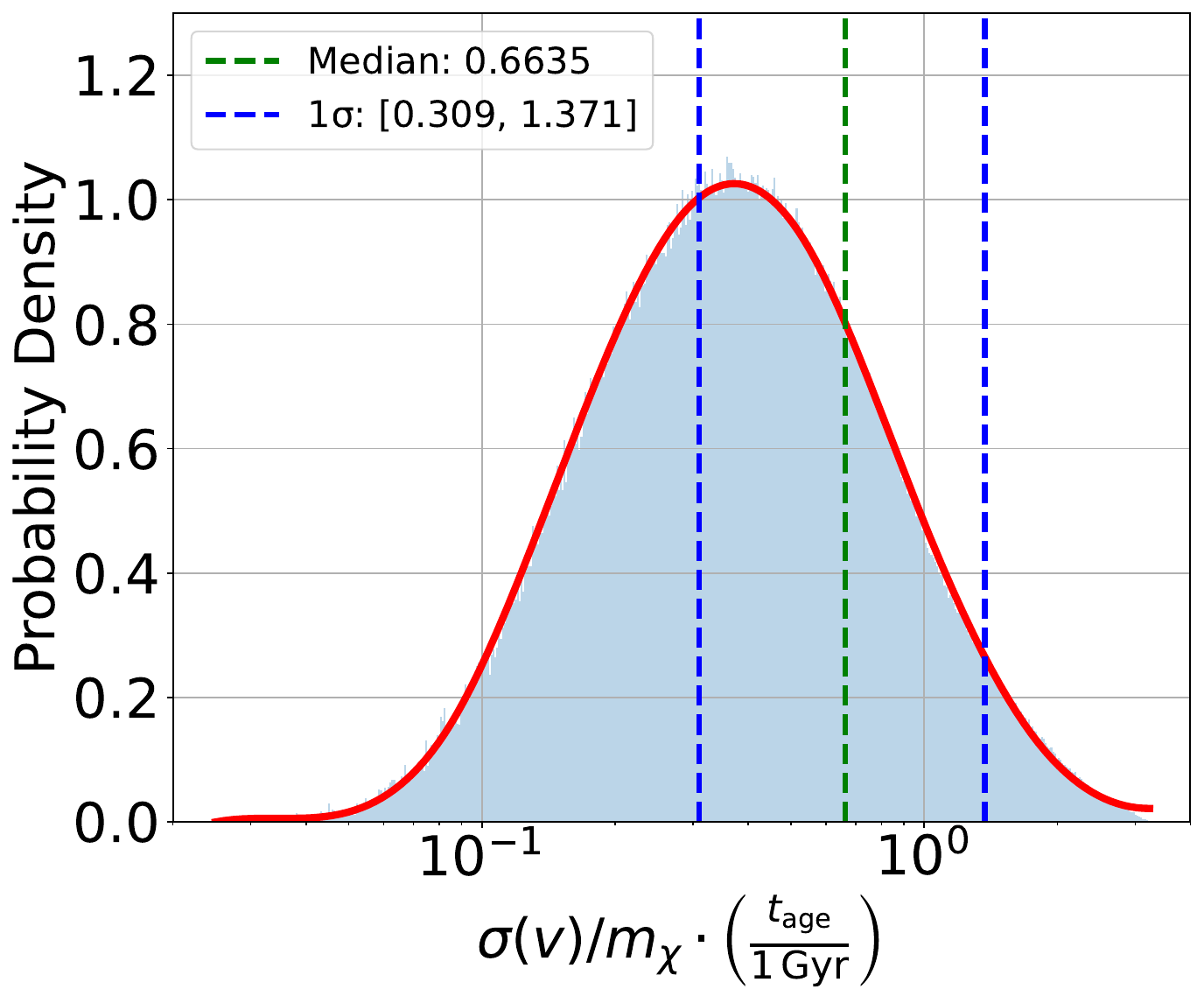}
    \vspace{-5pt}
    \caption{Upper: the posterior distribution of the interaction cross section at a relative dark matter particle velocity $v_{\mathrm{rel}} = 1500\,\mathrm{km/s}$, obtained by averaging over contributions from initial eccentricities. The green dashed line represents the central value, while the blue dashed lines indicate the $\pm1\sigma$ confidence level (CL) intervals. Lower: the posterior distribution of the interaction cross section at relative velocity $v_{\mathrm{rel}} = 1500\,\mathrm{km/s}$ obtained with fixed initial eccentricity $e_0 = 0.9$.}
    \label{fig:differ_e0_sigmavGyr_postdistribution}
\end{figure}

We present in Table~\ref{Tab:e0_parameter_distribution} the posterior estimates of the transition velocity and cross section constant under different initial eccentricity conditions. Physically we anticipate that both parameters should increase monotonically with eccentricity. The current data, however, are insufficient to substantiate this expectation—indeed, when the error bars are reduced, the estimated cross section decreases. This discrepancy may stem from the limited precision of the present data and from the differing spectral sensitivities and partial degeneracy between $v_t$ and $\sigma_0/m_{\chi}$. The inferred mediator to the dark matter mass ratio $m_{A'}/m_{\chi}=v_t$ grows monotonically with eccentricity, implying a range of $\left(m_{A'}/m_{\chi}\right)\lesssim 0.224\%$ for SIDM scenarios.

Meanwhile, in Fig.~\ref{fig:differ_e0_sigmavGyr_postdistribution}, we present the posterior distribution of the velocity-dependent cross section obtained from the posterior distributions of $v_t$ and $\sigma_0/m_{\chi}$, evaluated at a reference dark matter particle relative velocity of $v_{\mathrm{rel}}=1500\,\mathrm{km}/\mathrm{s}$. With a uniform initial eccentricity, averaging over all orbits gives $\sigma_0/m_{\chi}\simeq0.054\,\mathrm{cm^2\,g^{-1}}$ for a core age of $1\,\mathrm{Gyr}$. For highly eccentric binaries ($e_0 = 0.9$), the larger inferred $v_t$ raises the estimate to $\sigma(v)/m_{\chi}\simeq0.664\,\mathrm{cm^2\,g^{-1}}$, setting an upper bound for the SIDM model.

\begin{figure*}[t]
    \centering
    \includegraphics[width=0.8\textwidth]{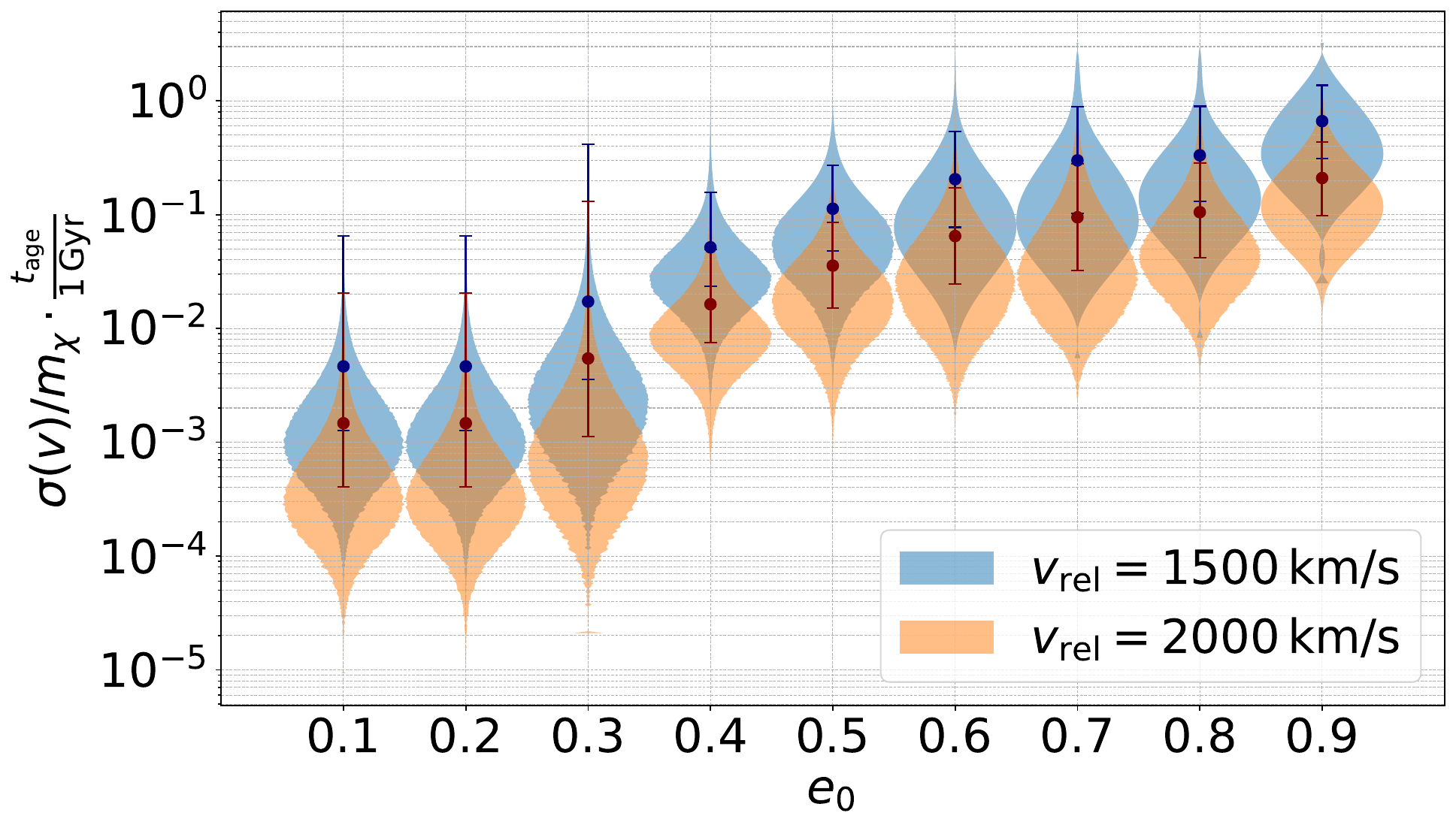}
    \vspace{5pt} 
    \includegraphics[width=0.8\textwidth]{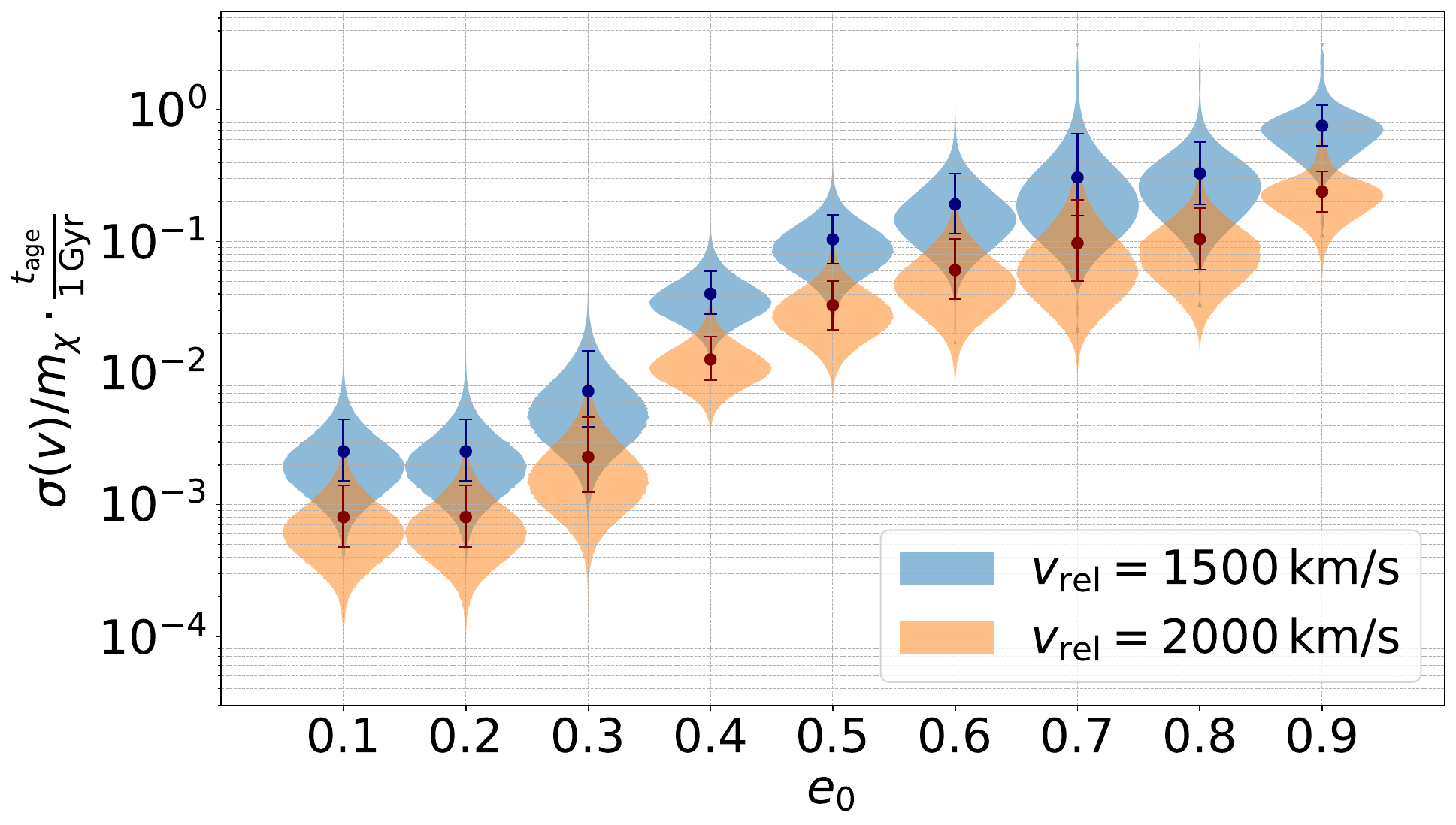}
    \vspace{-10pt}
    \caption{The upper figure shows violin plots of the posterior distribution of cross sections for dark matter particle relative velocities $v_{\mathrm{rel}}=1500$ and $v_{\mathrm{rel}}=2000\,\mathrm{km/s}$ under different fixed initial eccentricities. The error bars represent the central values and $\pm1\sigma$ CL intervals of the cross section posterior distributions. The lower figure displays violin plots of the projected cross section posterior distributions when the error of characteristic strain values from the PTA data are reduced by half in the future. Compared to the upper panel, the corresponding error bar intervals become narrower. Both plots clearly demonstrate the influence of initial eccentricity values on cross section estimation, and show that the estimated values increase monotonically with increasing initial eccentricity.}
    \label{fig:sigmav_e0_violin}
\end{figure*}

In Fig.~\ref{fig:sigmav_e0_violin}, we show the estimated range of $\sigma(v)/m_{\chi}$ at different initial eccentricities. We adopt a mean dark matter particle relative velocity of $1500 \,\mathrm{km}/\mathrm{s}$ and $2000 \,\mathrm{km}/\mathrm{s}$. Notably, the estimated cross section exhibits a monotonic increase with orbital eccentricity, a trend independent of whether initial eccentricity drives the orbit toward extreme values. 
\begin{comment}
    Our results demonstrate that, by including the most extreme contribution from source eccentricities to the SGWB, we obtain an upper limit on the cross section, $\sigma/m_{\chi}=0.664^{+0.708}_{-0.355}\,\mathrm{cm}^2/\mathrm{g}$ at the $68\%\mathrm{CL}$ for a reference velocity of $v_{\mathrm{rel}}=1500\,\mathrm{km\,s^{-1}}$.
\end{comment}
We also predict in the lower panel the distribution obtained when the error bars are halved in the future and the estimated range becomes noticeably narrower as the uncertainties decrease.

Together, these results have illustrated the impact of initial eccentricity on the cross section estimate and imply that by accounting for the additional contribution from eccentric orbits, an upper bound on the self-interacting cross section can be obtained. With improved precision in future PTAs data, we can thus expect narrower ranges for the models or the opportunity to test more refined theoretical scenarios.

Additionally, Table~\ref{compare_sigma_result} presents a comparative analysis of SIDM cross section limits derived under different reference velocities, benchmarked against existing observational and simulation-based studies. It shows we are able to offer new ranges for the cross section within corresponding velocity intervals for both dwarf galaxies and galaxy clusters. These ranges can also indicate upper limits on the interaction cross section derived under the framework of eccentric orbital dynamics. Furthermore, these results demonstrate the feasibility of utilizing PTA data to investigate the environment surrounding supermassive black holes. Particularly for various dark matter models, novel parameter ranges could be identified by quantifying deviations between the characteristic strain of the SGWB and model predictions.

\begin{table*}[t]
    \caption{Based on the fit ranges we obtained for the self-interacting cross section and transition velocity, we calculate the projected ranges of the cross section at particle velocities corresponding to the following astrophysical scales and compare them with previously established bounds.}
    \begin{ruledtabular}
        \renewcommand{\arraystretch}{1.8}
        \begin{tabular}{cccc}
            Observation & $\sigma/m_{\chi}$ & Our range: $\sigma/m_{\chi}$ & $\langle v_{\mathrm{rel}}\rangle$ \\
            \hline
            Rotation curves~\cite{Kaplinghat_2016} & $\geq 1\,\mathrm{cm}^2/\mathrm{g}$ & $1.048^{+0.65}_{-0.51}\,\mathrm{cm}^2/\mathrm{g}$ & $30-200\,\mathrm{km}/\mathrm{s}$ \\
            Stellar dispersion~\cite{Zavala_2013} & $\geq 0.6\,\mathrm{cm}^2/\mathrm{g}$ & $0.382^{+0.237}_{-0.188}\,\mathrm{cm}^2/\mathrm{g}$ & $50\,\mathrm{km}/\mathrm{s}$ \\
            \makecell{Stellar dispersion, \\ lensing~\cite{Elbert_2018, Kaplinghat_2016}} & $\sim 1\,\mathrm{cm}^2/\mathrm{g}$ &$0.664^{+0.708}_{-0.355}\,\mathrm{cm}^2/\mathrm{g}$ & $1500\,\mathrm{km}/\mathrm{s}$ \\
            \makecell{DM-galaxy offset \\ (Abell 3827)~\cite{Kahlhoefer_2015}} & $\sim 1.5\,\mathrm{cm}^2/\mathrm{g}$ &$\left(\frac{t_{\mathrm{age}}}{1\,\mathrm{Gyr}}\right)\cdot0.664^{+0.708}_{-0.355}\,\mathrm{cm}^2/\mathrm{g}$ & $1500\,\mathrm{km}/\mathrm{s}$ \\
            \makecell{DM-galaxy offset \\ (Abell 520)~\cite{Jee_2014, Kahlhoefer_2013, Spethmann_2017}} & $\sim 2\,\mathrm{cm}^2/\mathrm{g}$ &$\left(\frac{t_{\mathrm{age}}}{1\,\mathrm{Gyr}}\right)\cdot0.086^{+0.092}_{-0.046}\,\mathrm{cm}^2/\mathrm{g}$ & $\sim2500\,\mathrm{km}/\mathrm{s}$ \\
            \makecell{Cluster lensing surveys \\ (Ellipticity)~\cite{Peter_2013}} & $\leq 1\,\mathrm{cm}^2/\mathrm{g}$ &$\left(\frac{t_{\mathrm{age}}}{1\,\mathrm{Gyr}}\right)\cdot1.176^{+1.254}_{-0.628}\,\mathrm{cm}^2/\mathrm{g}$ & $1300\,\mathrm{km}/\mathrm{s}$ \\
            \makecell{DM-galaxy offset \\ (Substructure mergers)~\cite{Harvey_2015, Wittman_2018}} & $\lesssim 2\,\mathrm{cm}^2/\mathrm{g}$ & $\left(\frac{t_{\mathrm{age}}}{1\,\mathrm{Gyr}}\right)\cdot0.085^{+0.046}_{-0.029}\,\mathrm{cm}^2/\mathrm{g}$ & $500-4000\,\mathrm{km}/\mathrm{s}$ \\
            \makecell{Mass-to-light ratio \\ (Bullet Cluster)~\cite{Randall_2008}} & $< 0.7\,\mathrm{cm}^2/\mathrm{g}$ & $\left(\frac{t_{\mathrm{age}}}{1\,\mathrm{Gyr}}\right)\cdot0.013^{+0.014}_{-0.007}\,\mathrm{cm}^2/\mathrm{g}$ & $4000\,\mathrm{km}/\mathrm{s}$ \\  
        \end{tabular}
    \end{ruledtabular}
    \label{compare_sigma_result}
\end{table*}

\section{Conclusion}\label{sec:conclusion}

This work adopts an eccentric orbital SMBHB merger model surrounded by SIDM and incorporates additional drag effects in dynamical friction to calculate the binary's evolution and GW emission, thereby obtaining the corresponding SGWB spectrum. By comparing with the characteristic strain data of SGWB from NANOGrav, PPTA, and EPTA, we establish a Bayesian model and determine the current nHz-band GW detection's preferred range for SIDM model parameters.

Considering a uniform initial eccentricity distribution, the analysis yields preferred ranges for the mediator to dark matter mass ratio as $m_{A'}/m_\chi \approx 0.119^{+0.006}_{-0.004}\%$ and for the cross section as $\sigma(v)/m_\chi\cdot(t_{\mathrm{age}}/1\,\mathrm{Gyr})\approx0.054^{+0.251}_{-0.044}\,\mathrm{cm}^2/\mathrm{g}$ at $v_{\mathrm{rel}}=1500\,\mathrm{km/s}$. Furthermore, we demonstrate that the initial eccentricity influences the cross section estimation significantly, and our results suggest a preferred range of the mediator-to-dark matter mass ratio and the cross section such that $m_{A'}/m_{\chi}\lesssim0.22\%$ and $\sigma(v)/m_\chi\cdot(t_{\mathrm{age}}/1\,\mathrm{Gyr})\lesssim0.66\,\mathrm{cm}^2/\mathrm{g}$ at $v_{\mathrm{rel}}=1500\,\mathrm{km/s}$. These findings demonstrate the potential of SGWB observations to probe dark matter particle properties. We anticipate that the improvement in the precision of future PTAs data will lead to a more accurate estimation of dark matter model parameters.

\begin{acknowledgments}
This work is partly supported by the National Key Research and Development Program of China (Grant No. 2021YFC2201901), 
%the National Natural Science Foundation of China (Grant No.12347103) 
and the Fundamental Research Funds for the Central Universities. 
\end{acknowledgments}

\section*{DATA AVAILABILITY}
The data that support the findings of this article are not publicly available upon publication because it is not technically feasible and/or the cost of preparing, depositing, and hosting the data would be prohibitive within the terms of this research project. The data are available from the authors upon reasonable request.

\appendix
\section{POLYNOMIALS FOR NFW PARAMETERS}
\label{App:mass_ploynomials}
Equation~\eqref{poly1} describes the relation between the halo mass and the total stellar mass of galaxies using a double power-law distribution, where the polynomial can be expressed as~\cite{Girelli_2020}
\begin{align}
    \log M_A(z) &= (\log M_A)_{z=0} + z \cdot \mu = B + z \cdot \mu,  \\
    A(z) &= \left(\frac{M_*}{M_h}\right)_{z=0} \cdot (1 + z)^\nu = C \cdot (1 + z)^\nu,  \\
    \gamma(z) &= \gamma_0 \cdot (1 + z)^\eta = D \cdot (1 + z)^\eta,  \\
    \beta(z) &= F \cdot z + E, 
\end{align}
with the parameters of the polynomial we select being $B=11.83, \mu=0.18, C=0.047, \nu=-0.4, D=0.728, \eta=-0.16, F=0.052, E=0.92$.
For the polynomial in Eq.~\eqref{eq:poly2}, by fitting the data points from Table~\ref{poly2_parameter}, we can get the values with respect to redshift .
\begin{table}[t]
    \caption{The relationship between the values of the polynomial in Eq.~\eqref{eq:poly2} and the variation in redshift~\cite{2016MNRAS.457.4340K}.}
    \begin{ruledtabular}
        \begin{tabular}{cccc}
            Redshift & $C_c(z)$ & $\gamma_c(z)$ & $M_c(z) / 10^{12}h^{-1}M_\odot$ \\
            \hline
            0.00 & 7.40 & 0.120 & $5.5 \times 10^5$ \\
            0.35 & 6.25 & 0.117 & $1.0 \times 10^5$ \\
            0.50 & 5.65 & 0.115 & $2.0 \times 10^4$ \\
            1.00 & 4.30 & 0.110 & 900 \\
            1.44 & 3.53 & 0.095 & 300 \\
            2.15 & 2.70 & 0.085 & 42 \\
            2.50 & 2.42 & 0.080 & 17 \\
            2.90 & 2.20 & 0.080 & 8.5 \\
            4.10 & 1.92 & 0.080 & 2.0 \\
            5.40 & 1.65 & 0.080 & 0.3 \\
        \end{tabular}
    \end{ruledtabular}
    \label{poly2_parameter}
\end{table}

\section{ORBITAL DYNAMICS OF THE BINARY}
\label{App:ae_rate}
For a binary system with total mass $M_{\mathrm{tot}}$, component masses $m_1$ and $m_2$, and mass ratio $q=m_2/m_1$, where $q\in\left(0,1\right)$, the position of the $i$-th black hole in the binary system can be written as~\cite{Yue_2019, Li:2021pxf}
\begin{equation}
    r_i=\frac{a_i(1-e^2)}{1+e\mathrm{cos}(\phi_i)}
\end{equation}
where $a_i$ is the corresponding orbital semimajor axis. And the energy and angular momentum of each component are given as
\begin{align}
    E_i&=-\frac{Gm_iM_i}{2a_i},\\
    L_i&=m_i\sqrt{GM_ia_i(1-e^2)},
\end{align}
where $M_i$ satisfies $M_1 = \frac{m_2^3}{M_{\mathrm{tot}}^2}$ and $M_2 = \frac{m_1^3}{M_{\mathrm{tot}}^2}$, representing the effective central mass governing the orbital motion of each individual star. Then the total energy and total angular momentum can be written as
\begin{align}\label{aEL_relation}
    E&=E_1+E_2=-\frac{Gm_1m_2}{2a},\\
    L&=L_1+L_2=m_1m_2\sqrt{\frac{Ga(1-e^2)}{M_{\mathrm{tot}}}}.
\end{align}
The velocity and angular velocity of the components can be expressed using Kepler's laws as
\begin{align}
    v_i &= \sqrt{\frac{2}{m_i}\left(E_1+\frac{GM_im_i}{r_i}\right)}\\
    \dot{\phi_i}&=\frac{L_i}{m_ir^2_i}=\frac{\sqrt{GM_ia_i\left(1-e^2\right)}}{r^2_i}.
\end{align}
Then the specific form of Eq.~\eqref{eq:dE_dL} is as follows
\begin{align}
    \left\langle\frac{\mathrm{d}E_i}{\mathrm{d}t}\right\rangle_{\mathrm{DF}}&=\int_0^{2\pi}\mathrm{d}\phi_i \frac{2G^{\frac{3}{2}}m^2_i}{M^{\frac{1}{2}}_i}\left(\mathrm{log}\Lambda N_1 + N_2\right)\\ \nonumber
    &\quad\cdot\frac{r_i^{\frac{1}{2}}\rho_{\mathrm{sp}}(r_i)(1-e^2)^{\frac{3}{2}}}{(1+e^2+2e\mathrm{cos}\phi_i)^{\frac{1}{2}}(1+e\mathrm{cos}\phi_i)^{\frac{3}{2}}},\\
    \left\langle\frac{\mathrm{d}L_i}{\mathrm{d}t}\right\rangle_{\mathrm{DF}}&=\int_0^{2\pi}\mathrm{d}\phi_i \frac{2G m^2_i}{M_i}\left(\mathrm{log}\Lambda N_1 +N_2\right)\\ \nonumber
    &\quad\cdot\frac{r^2_i\rho_{\mathrm{sp}}(r_i)(1-e^2)^{\frac{3}{2}}}{(1+e^2+2e\mathrm{cos}\phi_i)^{\frac{3}{2}}},
\end{align}
where we use the relation $\frac{1}{T}\int_0^T\mathrm{d}t=\int_0^{2\pi}\frac{\mathrm{d}\phi}{2\pi}(1-e^2)^{\frac{3}{2}}(1+e\mathrm{cos}\phi)^{-2}$~\cite{10.1093/acprof:oso/9780198570745.001.0001}. We note that $N_1$ and $N_2$ are also functions of position, and their values are determined by Eq.~\eqref{eq:N1N2} mentioned earlier. Here we do not substitute the specific form of the density profile because for SIDM models, the power-law index is not constant near the transition radius. Therefore, it is necessary to obtain the density profile numerically before calculating the change rates of energy and angular momentum at each position. We sum the contributions of the two components and incorporate the effects of gravitational radiation. Utilizing Eq.~\eqref{aEL_relation} between the semimajor axis, eccentricity, and the energy and angular momentum, we can obtain
\begin{align}\label{eq:de_da}
    \dot{a} &= \frac{2a^2}{G m_1 m_2} \dot{E},\\
    \dot{e} &= \frac{a(1 - e^2)}{e G m_1 m_2} \dot{E} - \frac{\sqrt{M(1 - e^2)}}{e m_1 m_2 \sqrt{a G}} \dot{L}.
\end{align} 
By substituting these into Eq.~\eqref{eq:de_da}, the differential equation can be solved by using numerical methods.

\section{NUMBER DENSITY OF BLACK HOLE MERGERS}
\label{App:number_density}

\begin{table}[htb]
    \caption{Parameter values used in the model as~\cite{alon2024}.}
    \label{density_parameters}
    \begin{ruledtabular}
        \begin{tabular}{ll|ll}
            Parameter & Value & Parameter & Value \\
            \hline
            $\psi_0$ & $-2.31$ & $P_0$ & $0.033$ \\
            $\psi_z$ & $-0.6$ & $\beta_{p0}$ & $1$ \\
            $m_{\psi 0}$ & $11.5$ & $T_0$ & $0.5 \, \mathrm{Gyr}$ \\
            $m_{\psi z}$ & $0.11$ & $\beta_{t0}$ & $-0.5$ \\
            $\alpha_{\psi 0}$ & $-1.21$ & $\gamma_{t0}$ & $-1$ \\
            $\alpha_{\psi z}$ & $-0.03$ & & \\
        \end{tabular}
    \end{ruledtabular}
\end{table}

In Eq.~\eqref{event_density}, we convert the number density of supermassive black hole mergers into the merger event density of galaxies, and the polynomial in Eq.~\eqref{event_density_of_galaxy} can be expressed as~\cite{alon2024}
\begin{equation}
    \Psi(M_{\star},z)=\Psi_0\left(\frac{M_{\star}}{M_{\Psi}}\right)^{\alpha_{\Psi}}\mathrm{exp}\left(-\frac{M_{\star}}{M_{\Psi}}\right),
\end{equation}
where
\begin{align}
    \log_{10} \left( {\Psi_0}/{\mathrm{Mpc}^{-3}} \right) &= \psi_0 + \psi_z \cdot z,  \\ \label{eq:psi_constant_def}
    \log_{10} \left( {M_\Psi}/{M_\odot} \right) &= m_{\psi 0} + m_{\psi z} \cdot z,  \\
    \alpha_\Psi &= 1 + \alpha_{\psi 0} + \alpha_{\psi z} \cdot z. 
\end{align}
The galaxy pair function can be described by
\begin{equation}
    P\left(M_{\star},q_{\star},z\right)=P_0(1+z)^{\beta_{p0}},
\end{equation}
and the galaxy merger time
\begin{equation}
    T_{\mathrm{g}-\mathrm{g}}(M_{\star},q_{\star},z)=T_0(1+z)^{\beta_{t0}}q^{\gamma_{t0}}_{\star}.
\end{equation}
We have listed the values of the parameters in these polynomials in Table~\ref{density_parameters} as~\cite{alon2024}. We note that the redshift $z'$ used in these polynomials corresponds to the onset of galaxy mergers, and its relationship with the current redshift $z$ can be simply expressed as
\begin{align}
    t(z)-t(z')=T_{\mathrm{g}-\mathrm{g}}(z'),\;
    t(z=0)=13.79~\mathrm{Gyr},
\end{align}
where the relationship between redshift and time is described using the standard cosmological model,
\begin{align}
    \frac{\mathrm{d}t}{\mathrm{d}z}&=\frac{1}{(1+z)H(z)},\\
    H(z)&=H_0\left[\Omega_{\Lambda}+(1+z)^3\Omega_m\right]^{\frac{1}{2}},
\end{align}
and we take $H_0=67.4\mathrm{km}\mathrm{s}^{-1}\mathrm{Mpc}^{-1}, \Omega_m=0.315, \Omega_{\Lambda}=0.685$ here~\cite{2020}.

\begin{figure}[t]
    \centering
    \includegraphics[width=0.45\textwidth]{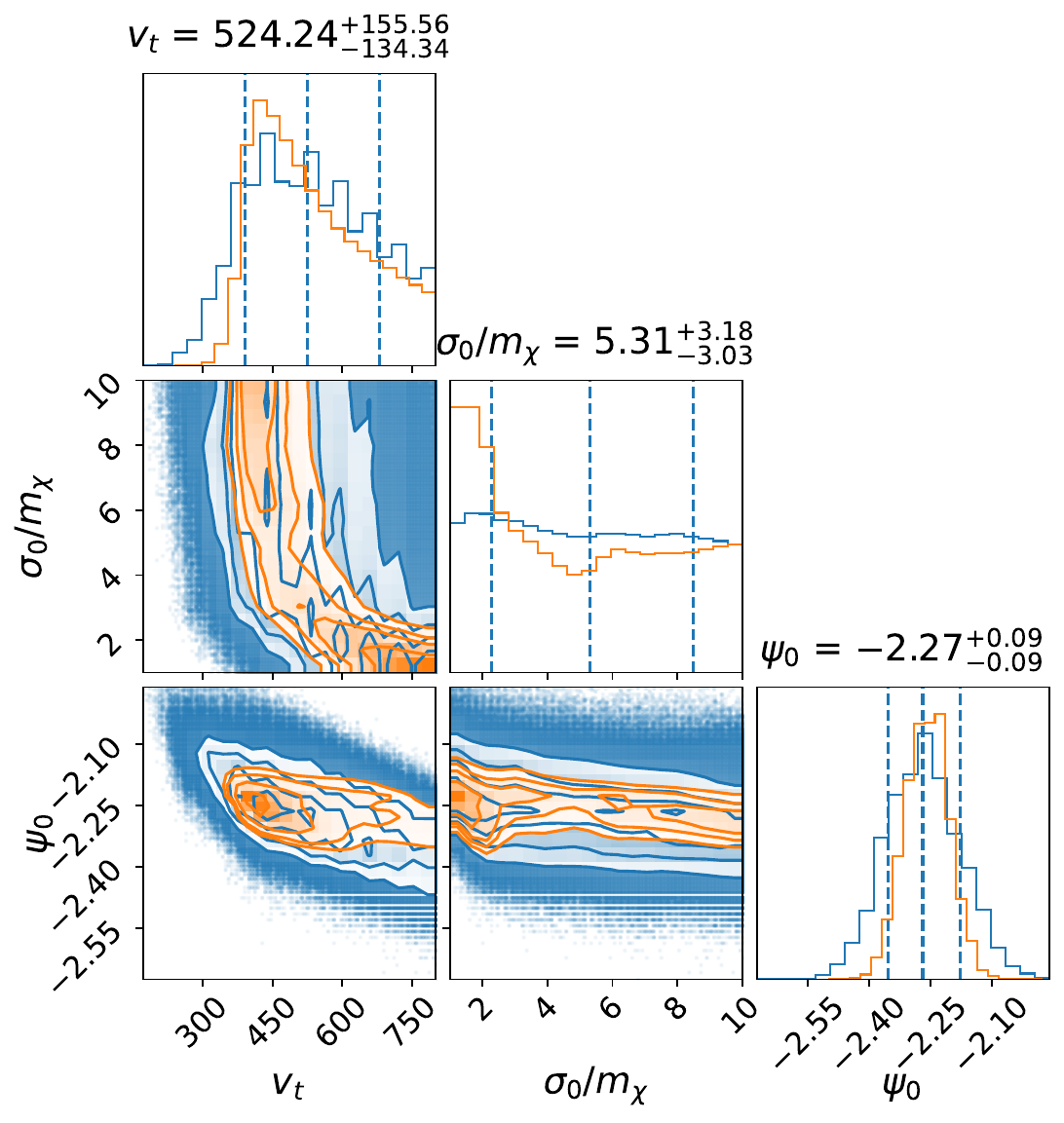} 
    \vspace{-5pt} 
    \includegraphics[width=0.45\textwidth]{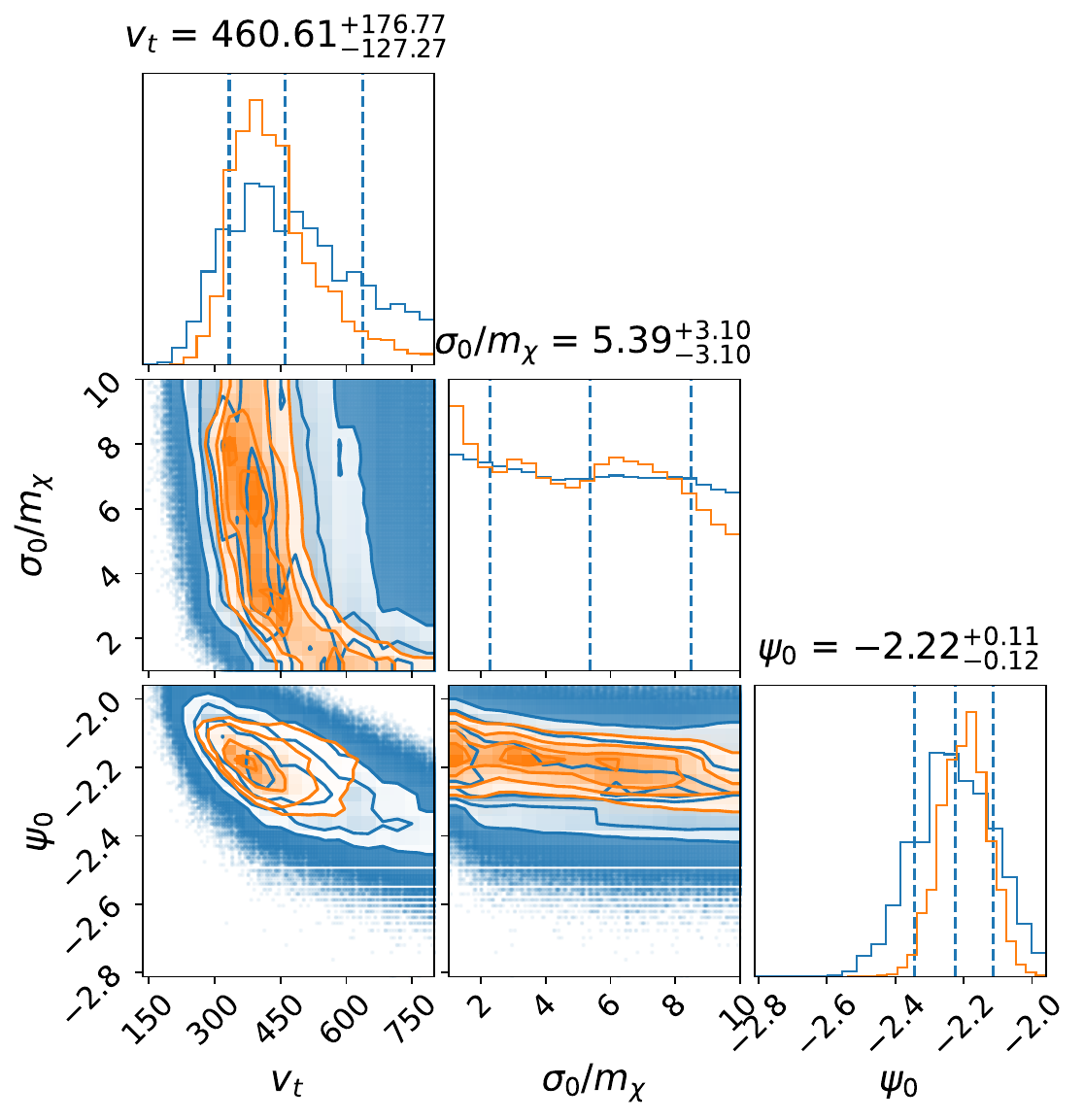}
    \vspace{-8pt}
    \caption{The posterior distributions of dark matter model parameters obtained when adopting an eccentricity distribution of $e_0=\mathcal{N}[0.7,0.3]$ indicate higher estimates for both the transition velocity and the cross section constant compared to those in Fig.~\ref{fig:result_paras}, along with a smaller normalization coefficient $\psi_0$.}
    \label{fig:postdistri_NormalD}
\end{figure}

\newpage
\bibliographystyle{apsrev4-2}
\bibliography{SGWB}

\end{document}